\documentclass[twoside,11pt]{article}
\usepackage{a4wide}
\usepackage{cite,latexsym,amsfonts,amssymb,exscale,epsfig,bbm}
\usepackage[centertags,sumlimits,intlimits,namelimits,reqno]{amsmath}
\usepackage{amsthm}
\theoremstyle{definition}
\newtheorem{theorem}{Theorem}[section]
\newtheorem{lemma}[theorem]{Lemma}
\newtheorem{corollary}[theorem]{Corollary}
\newtheorem{proposition}[theorem]{Proposition}
\newtheorem{definition}[theorem]{Definition}
\newtheorem{remark}[theorem]{Remark}
\newcommand{\ontop}[2]{\genfrac{}{}{0pt}{2}{\scriptstyle #1}{\scriptstyle #2}}
\def\nn{\notag}

\def\address#1{\date{{\sl #1}\\\ \\\theversion}\gdef\date##1{}}%
\def\version#1{\gdef\theversion{#1}}%

\def\dopreprint{\hfill{\small\thepreprint}\\}%
\def\preprint#1{\def\thepreprint{#1}}%
\def\thepreprint#1{}%

\def\sym#1{{\mathcal #1}}
\def\emph#1{{\sl #1\/}}

\let\phi=\varphi
\let\theta=\vartheta
\let\epsilon=\varepsilon

\def\SO{{SO}}

\def\SU{{SU}}

\def\Spin{{Spin}}

\def\dim{\mathop{\rm dim}\nolimits}
\def\Stab{\mathop{\rm Stab}\nolimits}

\def\Hom{\mathop{\rm Hom}\nolimits}
\def\Aut{\mathop{\rm Aut}\nolimits}

\def\openone{\mathbbm{1}}%
\def\pacs#1{\noindent PACS: #1\par}%
\def\keywords#1{\noindent key words: #1\par}%
\def\acknowledgements{\section*{Acknowledgements}}%

\def\mycaption#1#2{%
  \begin{quote}
  \caption{\label{#1}#2}
  \end{quote}}

\def\C{{\mathbbm C}}
\def\N{{\mathbbm N}}
\def\R{{\mathbbm R}}
\def\Z{{\mathbbm Z}}

\let\hat=\widehat
\let\tilde=\widetilde

\def\ie{{\sl i.e.\/}}

\def\etc{{\sl etc.\/}}

\def\Rep{\tilde{\sym{R}}}%
\def\Irrep{\sym{R}}%
\def\1{\mathbf{1}}
\def\Calg{C_{\rm alg}}%
\def\del{\partial}

\newcounter{mathletter}%
\newcommand{\bmathletter}{%
  \refstepcounter{equation}%
  \setcounter{mathletter}{\value{equation}}%
  \setcounter{equation}{0}%
    \renewcommand{\theequation}{%
      \mbox{\thesection.\arabic{mathletter}\alph{equation}}}}%
\newcommand{\emathletter}{\setcounter{equation}{\value{mathletter}}}%
\newenvironment{mathletters}{\bmathletter}{\emathletter}%

\makeatletter

\newfont{\@aidxte}{cmsy10}
\newfont{\@aidxel}{cmsy10 scaled 1095}
\newfont{\@aidxtw}{cmsy10 scaled 1200}
\newlength\@aidxtexvi
\newlength\@aidxtexvii
\newlength\@aidxelxvi
\newlength\@aidxelxvii
\newlength\@aidxtwxvi
\newlength\@aidxtwxvii
\newcommand{\alignidx}[1]{%
  \@aidxtexvi=\fontdimen16\@aidxte
  \@aidxtexvii=\fontdimen17\@aidxte
  \@aidxelxvi=\fontdimen16\@aidxel
  \@aidxelxvii=\fontdimen17\@aidxel
  \@aidxtwxvi=\fontdimen16\@aidxtw
  \@aidxtwxvii=\fontdimen17\@aidxtw
    {\mbox{$%
    \fontdimen16\@aidxte=2.9pt
    \fontdimen17\@aidxte=2.9pt
    \fontdimen16\@aidxel=3.1pt
    \fontdimen17\@aidxel=3.1pt
    \fontdimen16\@aidxtw=3.3pt
    \fontdimen17\@aidxtw=3.3pt
    #1$}}%
    \fontdimen16\@aidxte=\@aidxtexvi
    \fontdimen17\@aidxte=\@aidxtexvii
    \fontdimen16\@aidxel=\@aidxelxvi
    \fontdimen17\@aidxel=\@aidxelxvii
    \fontdimen16\@aidxtw=\@aidxtwxvi
    \fontdimen17\@aidxtw=\@aidxtwxvii}

\renewcommand{\theequation}{\thesection.\arabic{equation}}
\@addtoreset{equation}{section}

\makeatother

\newenvironment{myenumerate}{%
  \begin{enumerate}
  \setlength{\partopsep}{0pt}
  \setlength{\parskip}{0pt}}{\end{enumerate}}

\version{30 January 2002}%
\preprint{DAMTP-2001-103}

\begin{document}
%

\title{\dopreprint Dual variables and a connection picture for the
        Euclidean Barrett--Crane model} 
\author{Hendryk Pfeiffer\thanks{e-mail: H.Pfeiffer@damtp.cam.ac.uk}}
\address{Emmanuel College, St.~Andrew's Street, Cambridge CB2 3AP\\
         and\\
         Department of Applied Mathematics and Theoretical Physics,\\
         Wilberforce Road, Cambridge CB3 0WA\\
         England, UK}
\date{\version}
\maketitle

%
\begin{abstract}
%

  The partition function of the $\SO(4)$- or $\Spin(4)$-symmetric
  Euclidean Barrett--Crane model can be understood as a sum over all
  quantized geometries of a given triangulation of a four-manifold. In
  the original formulation, the variables of the model are balanced
  representations of $\SO(4)$ which describe the quantized areas of
  the triangles. We present an exact duality transformation for the
  full quantum theory and reformulate the model in terms of new
  variables which can be understood as variables conjugate to the
  quantized areas. The new variables are pairs of $S^3$-values
  associated to the tetrahedra. These $S^3$-variables parameterize the
  hyperplanes spanned by the tetrahedra (locally embedded in $\R^4$),
  and the fact that there is a pair of variables for each tetrahedron
  can be viewed as a consequence of an $\SO(4)$-valued parallel
  transport along the edges dual to the tetrahedra. We reconstruct the
  parallel transport of which only the action of $\SO(4)$ on $S^3$ is
  physically relevant and rewrite the Barrett--Crane model as an
  $\SO(4)$ lattice $BF$-theory living on the $2$-complex dual to the
  triangulation subject to suitable constraints whose form we derive
  at the quantum level. Our reformulation of the Barrett--Crane model
  in terms of continuous variables is suitable for the application of
  various analytical and numerical techniques familiar from
  Statistical Mechanics.
\end{abstract}


%
\section{Introduction}
%

Spin foam models have been proposed as candidates for a quantum theory
of space-time geometry~\cite{Ba94,Re94,ReRo97,BaCr98,Ba98a} and
therefore form candidates for a quantum theory of gravity. Spin foam
models also play a central role in the construction of invariants of
piecewise-linear three- and four-manifolds~\cite{TuVi92,CrKa97}. They
arise as the models that are strong-weak dual to non-Abelian lattice
gauge theory~\cite{OePf01,PfOe02} and form a possible starting point
for the generalization of gauge theories from Lie groups to quantum
groups~\cite{Pf01,Oe02}. For reviews on the subject, see, for example
\cite{Ba99,Or01a}.

A spin foam~\cite{Ba98a} whose symmetry group is a compact Lie group
$G$, is an abstract oriented $2$-complex consisting of faces, edges
and vertices, together with a colouring of the faces with
representations of $G$ and a colouring of the edges with compatible
intertwiners (representation morphisms) of $G$. A spin foam model is a
discrete model in the framework of Statistical Mechanics whose
partition function is a sum over spin foams each of which is assigned
a Boltzmann weight. If the Boltzmann weight factors into contributions
for each vertex, edge and face, these factors are usually called
vertex, edge and face amplitudes.

Barrett, Crane and Baez~\cite{BaCr98,Ba98a,BaBa00} have developed a
particular spin foam model whose partition function can be understood
as the sum over all quantized Euclidean geometries that can be
assigned to a given triangulation of a four-manifold. The spin foam
model is usually defined on the $2$-complex dual to that
triangulation. Several versions of this model are studied in the
literature~\cite{DPFr00,PeRo01,OrWi01} which differ in their Boltzmann
weights (amplitudes), in particular in their edge amplitudes. Some of
the models are formulated on a fixed $2$-complex while others include
a sum over $2$-complexes. In the following, we call all these models
Barrett--Crane models.

In this article, we study a generalized version which encompasses all
interesting choices of amplitudes, but we restrict ourselves to a
fixed $2$-complex. The partition function of our model is a sum over
all possible assignments of \emph{balanced} irreducible
representations of the Lie group $\SO(4)$ to the faces. The balanced
representations are of the form $V\otimes V$ where $V$ is an
irreducible representation of $\SU(2)$. They are also called the
\emph{simple} representations. Each edge is then assigned a unique
intertwiner, the Barrett--Crane intertwiner~\cite{BaCr98}. We allow
quite generic edge and face amplitudes, but the vertex amplitude is
always the generalized $10j$-symbol~\cite{BaCr98,Ba98a} consisting of
one Barrett--Crane intertwiner for each edge attached to the vertex.

We make use of the fact that the Barrett--Crane intertwiner can be
written in terms of an integral over the sphere
$S^3$~\cite{Ba98b,FrKr00} and of the techniques familiar from the
duality transformation for non-Abelian lattice gauge
theory~\cite{OePf01,Pf01} and derive a dual expression for the
partition function of the Barrett--Crane model in which the
fundamental variables are pairs of $S^3$-values associated to the
edges. The partition sum is then given by an integral over $S^3$ for
each of these variables.

Provided that the edge amplitude of the original model is of a
particular form, the dual model has local interactions, its Boltzmann
weight factors into one contribution for each face, and it depends on
the $S^3$-variables at the edges in the boundary of that face.

This reformulation of the Barrett--Crane model has the following
geometric interpretation if the $2$-complex is chosen to be dual to a
triangulation of a four-manifold. The $S^3$-variables
are then associated to the tetrahedra. They parameterize the
(normalized) normal vectors of the hyperplanes spanned by the
tetrahedra where the tetrahedra are locally embedded in $\R^4$. The
factorized Boltzmann weight provides an interaction term for each
triangle so that the interaction depends on the $S^3$-values
associated to all tetrahedra that contain the triangle in their
boundaries.

However, there are two $S^3$-values associated to each tetrahedron one
of which belongs to either one of the two four-simplices that contain
the tetrahedron in their boundaries. This suggests that there exists
an $\SO(4)$-valued parallel transport along the tetrahedron which maps
the first $S^3$-value to the second one. However, this parallel
transport is defined only up to elements of $\SO(4)$ that stabilize
the first of the $S^3$-values. Following this idea, we show that the
Barrett--Crane model can be obtained from $\SO(4)$ lattice $BF$-theory
(the $\SO(4)$ Ooguri model~\cite{Oo92,KaSa00}) if we impose a
constraint at the quantum level by averaging over this stabilizer in a
suitable way. This means that of the $\SO(4)$-valued parallel
transport only the action of $\SO(4)$ on $S^3$ is physically relevant.

Finally, we consider the Boltzmann weight of the dual formulation of
the Barrett--Crane model and list possible face and edge amplitudes
for which the partition function is well defined (not just
divergent). We outline how one can analyze this model in the context
of Statistical Mechanics, for example in order to study its ground
state and fluctuations around it. In addition, one can expect that the
dual formulation of the Barrett--Crane model is well suited for
numerical studies.

The present article is organized as follows. In
Section~\ref{sect_prelim}, we review some mathematical background
material on the algebra of representation functions of a compact Lie
group, we construct the spherical functions of $S^3$, and we introduce
our notation for abstract $2$-complexes. The Barrett--Crane model and
$\SO(4)$ lattice $BF$-theory are introduced in
Section~\ref{sect_bcmodel}. In Section~\ref{sect_transform}, we
present the duality transformation for the Barrett--Crane model in
detail and show how this model is related to $\SO(4)$ lattice $BF$-theory
subject to a suitable constraint. We also discuss its geometric
interpretation and the possible choices of face and edge
amplitudes. Section~\ref{sect_conclusion} finally contains our
concluding comments.

Note added: The Barrett--Crane model is often understood as a path
integral in the `real time' picture, \ie\ using an integrand of the
form $\exp(iS)$ for some action $S$. However, Baez and
Christensen~\cite{BaCh01} have shown that for the common choices of
edge amplitudes, the integrand of the partition function is always
positive so that the path integral admits an interpretation in the
framework of Statistical Mechanics in terms of probabilities. In this
article, we use the language of Statistical Mechanics.

%
\section{Preliminaries}
%
\label{sect_prelim}

In this section, we briefly summarize definitions and some basic
statements related to the algebra of representation functions
$\Calg(G)$ of $G$ where $G$ is a compact Lie group. Furthermore, we
describe the spherical functions of $S^3$ and present our notation for
the $\SO(4)$ and $\SU(2)\times\SU(2)$ actions on $S^3$. More
background material and most of the proofs can be found, for example,
in~\cite{ViKl93,CaSe95}. Finally, we introduce the $2$-complexes that
we employ in our definition of the Barrett--Crane model and in the
duality transformation.

\subsection{The Hopf algebra of representation functions}

Let $G$ be a compact Lie group. We denote finite-dimensional complex
vector spaces on which $G$ is represented by $V_\rho$ and by
$\rho\colon G\to\Aut V_\rho$ the corresponding group
homomorphisms. Since each finite-dimensional complex representation of
$G$ is equivalent to a unitary representation, we select a set $\Rep$
containing one unitary representation of $G$ for each equivalence
class of finite-dimensional representations. The tensor product, the
direct sum and taking the dual are supposed to be closed operations on
this set. This amounts to a particular choice of representation
isomorphisms $\rho_1\otimes\rho_2\leftrightarrow\rho_3$
\etc, $\rho_j\in\Rep$, which is implicit in our formulas. We
furthermore denote by $\Irrep\subseteq\Rep$ the subset of irreducible
representations.

For a representation $\rho\in\Rep$, the dual representation is denoted
by $\rho^\ast$, and the dual vector space of $V_\rho$ by
$V_\rho^\ast$. The dual representation is given by $\rho^\ast\colon
G\mapsto \Aut V_\rho^\ast$, where
\begin{equation}
\label{eq_dualrep}
  \rho^\ast(g)\colon V_\rho^\ast\to V_\rho^\ast,\quad
    \eta\mapsto \eta\circ\rho(g^{-1}),
\end{equation}
\ie\ $(\rho^\ast(g)\eta)(v)=\eta(\rho(g^{-1})v)$ for all $v\in
V_\rho$. There exists a one-dimensional `trivial' representation of
$G$ which is denoted by $V_{[1]}\cong\C$.

For the unitary representations $V_\rho$, $\rho\in\Rep$, we have
standard (sesquilinear) scalar products $\left<\cdot;\cdot\right>$ and
orthonormal bases ${\{v_j\}}_j$. For each basis ${\{v_j\}}_j$ of
$V_\rho$, there exists a dual basis ${\{\eta^j\}}_j$ of $V_\rho^\ast$
so that $\eta^j(v_k)=\left<v_j;v_k\right>$, and $V_\rho^\ast$
possesses a scalar product such that
$\bigl<\eta^j;\eta^k\bigr>=\eta^k(v_j)$, $1\leq j,k\leq\dim V_\rho$.

The complex-valued functions
\begin{equation}
  t_{\eta,v}^{(\rho)}\colon G\to\C,\quad g\mapsto\eta(\rho(g)v),
\end{equation}
where $\rho\in\Rep$, $v\in V_\rho$ and $\eta\in V_\rho^\ast$, are
called the \emph{representation functions} of $G$. They form a
commutative and associative unital algebra over $\C$,
\begin{equation}
  \Calg(G) := \{\,t_{\eta,v}^{(\rho)}\colon\quad
    \rho\in\Rep, v\in V_\rho, \eta\in V_\rho^\ast\,\},
\end{equation}
whose operations are given by
\begin{mathletters}
\begin{eqnarray}
\label{eq_operationsum}
  (t_{\eta,v}^{(\rho)} + t_{\theta,w}^{(\sigma)})(g) &:=&
    t_{\eta+\theta,v+w}^{(\rho\oplus\sigma)}(g),\\
\label{eq_operationprod}
  (t_{\eta,v}^{(\rho)}\cdot t_{\theta,w}^{(\sigma)})(g)
    &:=& t_{\eta\otimes\theta,v\otimes w}^{(\rho\otimes\sigma)}(g),
\end{eqnarray}%
\end{mathletters}%
where $\rho,\sigma\in\Rep$, $v\in V_\rho$, $w\in V_\sigma$, $\eta\in
V_\rho^\ast$, $\theta\in V_\sigma^\ast$ and $g\in G$. The zero element
of $\Calg(G)$ is given by $t_{0,0}^{[1]}(g)=0$ and its unit element by
$t_{\eta,v}^{[1]}(g)=1$.

The algebra $\Calg(G)$ is furthermore equipped with a Hopf algebra
structure employing the coproduct
$\Delta\colon\Calg(G)\to\Calg(G)\otimes\Calg(G)\cong\Calg(G\times G)$,
the co-unit $\epsilon\colon\Calg(G)\to\C$ and the antipode
$S\colon\Calg(G)\to\Calg(G)$ which are defined by
\begin{mathletters}
\begin{eqnarray}
  \Delta t_{\eta,v}^{(\rho)} (g,h) 
    &:=& t_{\eta,v}^{(\rho)}(g\cdot h),\\
  \epsilon t_{\eta,v}^{(\rho)}
    &:=& t_{\eta,v}^{(\rho)}(e),\\
  S t_{\eta,v}^{(\rho)} (g) 
    &:=& t_{\eta,v}^{(\rho)}(g^{-1}),
\end{eqnarray}%
\end{mathletters}%
for $\rho\in\Rep$, $v\in V_\rho$, $\eta\in V_\rho^\ast$ and $g,h\in
G$. Here $e\in G$ denotes the group unit.

In the standard orthonormal bases, the representation functions are
given by the coefficients of representation matrices,
\begin{equation}
  t_{mn}^{(\rho)}(g) = \eta^m(\rho(g)v_n) =
  \left<v_m;\rho(g)v_n\right> = {\rho(g)}_{mn},
\end{equation}
so that the coproduct corresponds to the matrix product,
\begin{equation}
\label{eq_matrixcopro}
  \Delta t_{mn}^{(\rho)}(g,h) 
    = \sum_{j=1}^{\dim V_\rho} t_{mj}^{(\rho)}(g)t_{jn}^{(\rho)}(h)
    = \sum_{j=1}^{\dim V_\rho} {\rho(g)}_{mj}{\rho(h)}_{jn},
\end{equation}
while the antipode refers to the inverse matrix, $S
t_{mn}^{(\rho)}(g)={({\rho(g)}^{-1})}_{mn}$, and the co-unit describes
the coefficients of the unit matrix, $\epsilon
t_{mn}^{(\rho)}=\delta_{mn}$. Furthermore, the antipode relates a
representation to its dual,
\begin{eqnarray}
\label{eq_antipode}
  S t_{mn}^{(\rho)}(g) = \eta^m({\rho(g)}^{-1}v_n) 
    = (\rho^\ast(g)\eta^m)(v_n)
    = \left<\eta^n;\rho^\ast(g)\eta^m\right> 
    = t_{nm}^{(\rho^\ast)}(g),
\end{eqnarray}
which, in the case of unitary $\rho$, is just the conjugate
representation because on the other hand
$St_{mn}^{(\rho)}(g)=\overline{{\rho(g)}_{nm}}=\overline{t_{nm}^{(\rho)}(g)}$.
Here the bar denotes complex conjugation.

\subsection{Peter--Weyl decomposition and theorem}

The structure of the algebra $\Calg(G)$ can be understood if
$\Calg(G)$ is considered as a representation of $G\times G$ by
combined left and right translation of the function argument,
\begin{equation}
  (G\times G)\times\Calg(G)\to\Calg(G),\quad
    ((g_1,g_2),f)\mapsto (h\mapsto f(g_1^{-1}hg_2)).
\end{equation}
It can then be decomposed into its irreducible components as a
representation of $G\times G$.

\begin{theorem}[Peter--Weyl decomposition]
\label{thm_peterweyl}
Let $G$ be a compact Lie group.
\begin{myenumerate}
\item 
  There is an isomorphism
\begin{equation}
\label{eq_structure_calg}
  \Calg(G)\cong
    \bigoplus_{\rho\in\Irrep}(V_\rho\otimes V_\rho^\ast),
\end{equation}
  of representations of $G\times G$. Here the direct sum runs over the
  equivalence classes of finite-dimensional irreducible
  representations of $G$. The direct summands $V_\rho\otimes
  V_\rho^\ast$ are irreducible as representations of $G\times G$.
\item
  The direct sum in~\eqref{eq_structure_calg} is orthogonal with
  respect to the $L^2$-scalar product on $\Calg(G)$ which is formed
  using the Haar measure of $G$ on the left hand side, and using the
  standard scalar products on the right hand side, namely
\begin{equation}
\label{eq_l2measure}
  {\bigl<t_{\eta,v}^{(\rho)};t_{\theta,w}^{(\sigma)}\bigr>}_{L^2}
    := \int_G\overline{t_{\eta,v}^{(\rho)}(g)}\cdot t_{\theta,w}^{(\sigma)}(g)\,dg
    = \frac{1}{\dim V_{\rho}}\delta_{\rho\sigma}
        \overline{\left<\eta;\theta\right>}\left<v;w\right>,
\end{equation}
  where $\rho,\sigma\in\Irrep$ are irreducible. The Haar measure is
  denoted by $\int_G\,dg$ and normalized so that $\int_G\,dg=1$.
\end{myenumerate}
\end{theorem}

\begin{corollary}
Each representation function $f\in\Calg(G)$ can be decomposed
according to~\eqref{eq_structure_calg},
\begin{equation}
\label{eq_peterweyl_series}
  f(g) = \sum_{\rho\in\Irrep}\sum_{m,n=1}^{\dim V_\rho}
    \hat f^{(\rho)}_{mn}\,t^{(\rho)}_{mn}(g),\quad\mbox{where}\quad
  \hat f^{(\rho)}_{mn}=\dim V_\rho\,\int_G\overline{t^{(\rho)}_{mn}(g)}f(g)\,dg.  
\end{equation}
\end{corollary}

\noindent
The analytical aspects of $\Calg(G)$ can now be stated.

\begin{theorem}[Peter--Weyl Theorem]
Let $G$ be a compact Lie group. Then $\Calg(G)$ is dense in $L^2(G)$
with respect to the $L^2$ norm.
\end{theorem}

We use the Peter--Weyl theorem in order to complete $\Calg(G)$ with
respect to the $L^2$ norm to $L^2(G)$. Functions $f\in L^2(G)$ then
correspond to square summable series
in~\eqref{eq_peterweyl_series}. These series are therefore invariant
under a reordering of summands, and their limits commute with group
integrations. We make use of these invariances in the duality
transformation.

\subsection{Character decomposition}

The \emph{characters} of $G$ are the algebraic class functions, \ie\
those functions $f\in\Calg(G)$ that satisfy $f(hgh^{-1})=f(g)$ for all
$g,h\in G$.

\begin{proposition}
For class functions $f\in\Calg(G)$, the Peter--Weyl
decomposition~\eqref{eq_peterweyl_series} specializes to the
\emph{character decomposition}
\begin{equation}
\label{eq_charexp}
  f(g) = \sum_{\rho\in\Irrep}\chi^{(\rho)}(g)\,\hat f_\rho,
    \qquad\mbox{where}\qquad 
  \hat f_\rho = \int_G\overline{\chi^{(\rho)}(g)}f(g)\,dg.
\end{equation}
Here
\begin{equation}
  \chi^{(\rho)}:=\sum_{n=1}^{\dim V_\rho}t^{(\rho)}_{nn}
\end{equation}
denotes the character of the representation $\rho\in\Rep$. For
irreducible $\rho,\sigma\in\Irrep$, the orthogonality
relation~\eqref{eq_l2measure} implies,
\begin{equation}
  \bigl<\chi^{(\rho)};\chi^{(\sigma)}\bigr>_{L^2}
  = \int_G\overline{\chi^{(\rho)}(g)}\chi^{(\sigma)}(g)\,dg
  = \delta_{\rho\sigma}.
\end{equation}
\end{proposition}

The formal $\delta$-element,
\begin{equation}
  \delta(g) := \sum_{\rho\in\Irrep}\dim V_\rho\cdot\chi^{(\rho)}(g),
\end{equation}
satisfies for all $f\in\Calg(G)$,
\begin{equation}
  \int_G f(g)\delta(gh^{-1})\,dg=f(h).
\end{equation}

\subsection{Diagrams}

\begin{figure}[t]
\begin{center}
\input{pstex/diagrams.pstex_t}
\end{center}
\mycaption{fig_diagrams}{%
  Diagrams to visualize the index structure in the calculation of
  group integrals. (a) The identity map of $V_\rho$; (b) a
  representation function $t^{(\rho)}_{mn}$; (c) a product of
  representation functions $t^{(\rho_1)}_{m_1n_1}\cdots
  t^{(\rho_r)}_{m_rn_r}$; (d) the Haar intertwiner.}
\end{figure}

In order to present the partition function of lattice gauge theory
both in the original formulation using group integrations and in the
dual formulation~\cite{OePf01,Pf01} as a spin foam model, we define a
particular intertwiner~\cite{Oe02} based on the Haar measure.

\begin{definition}
Let $G$ be a compact Lie group and $\rho_1,\ldots,\rho_r\in\Irrep$,
$r\in\N$, be finite-dimensional irreducible representations of
$G$. The \emph{Haar intertwiner},
\begin{mathletters}
\label{eq_haarinter}
\begin{equation}
  T\colon\bigotimes_{\ell=1}^rV_{\rho_\ell}\to\bigotimes_{\ell=1}^rV_{\rho_\ell},
\end{equation}%
is defined by its matrix elements,
\begin{equation}
\label{eq_haarintermatrix}
  T_{m_1m_2\ldots m_r;n_1n_2\ldots n_r} := \int_G
    t^{(\rho_1)}_{m_1n_1}(g)t^{(\rho_2)}_{m_2n_2}(g)\cdots t^{(\rho_r)}_{m_rn_r}(g)\,dg.
\end{equation}%
\end{mathletters}%
\end{definition}

\begin{lemma}
The Haar intertwiner $T$ of~\eqref{eq_haarinter} satisfies for all
$h\in G$,
\begin{eqnarray}
  T &=& \int_G(\rho_1(g)\otimes\cdots\otimes\rho_r(g))\,dg,\\
  T &=& (\rho_1(h)\otimes\cdots\otimes\rho_r(h))\circ T 
     = T\circ (\rho_1(h)\otimes\cdots\otimes\rho_r(h)),\\
\label{eq_interproj}
  T\circ T&=&T  
\end{eqnarray}
The first equation is just the definition while the second and third
are consequences of the left-right translation invariance of the Haar
measure. In particular, $T$ forms a morphism of representations of $G$.
\end{lemma}

In the subsequent sections, the Haar intertwiner will appear in rather
complicated calculations. It is therefore convenient to introduce
diagrams that visualize the structure of the indices in these formulas
(Figure~\ref{fig_diagrams}).

The diagrams are read from top to bottom. We draw directed lines which
are labelled with irreducible representations $\rho\in\Irrep$ of
$G$. If the arrow points down, the line denotes the identity map of
$V_\rho$, Figure~\ref{fig_diagrams}(a). If the arrow points up, it
refers to the identity map of the dual representation $V_\rho^\ast$. A
representation function $t^{(\rho)}_{mn}$ is denoted by a box with an
incoming and an outgoing line (b), and a product of representation
functions by boxes placed next to each other (c). The Haar intertwiner
is visualized by the box labelled $T$ in (d).

\subsection{Spherical functions of $S^3$}
\label{sect_s3}

We construct the spherical functions of $S^3$ from the identification
$S^3\cong\SU(2)$, \ie\ we can make use of the representations
functions of $\SU(2)$. The finite-dimensional irreducible
representations of $\SU(2)$ are characterized by half-integers
$j\in\frac{1}{2}\N_0$. We write $V_j$, $\dim V_j=2j+1$, for the
irreducible representations. Using the Peter--Weyl
decomposition~\eqref{eq_peterweyl_series}, we can give a basis for
$\Calg(\SU(2))$,
\begin{equation}
  \{\,t^{(j)}_{mn}\colon\SU(2)\to\C\colon\quad j\in\frac{1}{2}\N_0,\quad
    1\leq m,n\leq2j+1\,\}.
\end{equation}

In particular, we obtain the (algebraic) spherical functions of $S^3$
from the identification $S^3\cong\SU(2)$ as
\begin{equation}
\label{eq_spherical}
  H^{(j)}_{mn}\colon S^3\to\C, g\mapsto t^{(j)}_{mn}(g),
\end{equation}
for $j\in\frac{1}{2}\N_0$ and $1\leq m,n\leq2j+1$.

We also wish to describe the transitive actions of $\SO(4)$ and
$\Spin(4)\cong\SU(2)\times\SU(2)$ on $S^3$ for which the stabilizer of
a point $g\in S^3$ is $\SO(3)$ or $\SU(2)$, respectively. Writing
$2\times 2$ complex matrices with the usual properties for the
elements of $\SU(2)$, we obtain $\SO(3)$ as the quotient
$\SO(3):=\SU(2)/\Z_2$ where $\Z_2:=\{-\openone,\openone\}\leq\SU(2)$
and $\SO(4):=(\SU(2)\times\SU(2))/\Z_2$ where
$\Z_2:=\{(\openone,\openone),(-\openone,-\openone)\}\leq\SU(2)\times\SU(2)$.
The relevant actions are the following.

\begin{proposition}
\label{prop_su2action}
The map
\begin{equation}
  (\SU(2)\times\SU(2))\times S^3\to S^3,((g_1,g_2),g)\mapsto
  (g_1,g_2)\ast g:=\alignidx{g_1gg_2^{-1}},
\end{equation}
is a transitive group action. The stabilizer of a point $g\in S^3$ is
given by
\begin{equation}
  \Stab_{\SU(2)\times\SU(2)}(g):=\{\,(ghg^{-1},h)\colon\quad
  h\in\SU(2)\,\}, 
\end{equation}
and forms a subgroup of $\SU(2)\times\SU(2)$ via the inclusion
\begin{equation}
  \imath^\prime_g\colon\SU(2)\hookrightarrow\SU(2)\times\SU(2),h\mapsto
  (ghg^{-1},h).
\end{equation}
The action $\ast$ on $S^3$ is well defined for the quotient
$\SO(4)\cong(\SU(2)\times\SU(2))/\Z_2$ in which case the stabilizer is
isomorphic to $\SO(3)$ and given by the inclusion
\begin{equation}
  \imath_g\colon\SO(3)\hookrightarrow\SO(4),h\mapsto (ghg^{-1},h),
\end{equation}
which we have written down for representatives.
\end{proposition}

\begin{corollary}
\label{corr_decomposes3}
The algebra of spherical functions $\Calg(S^3)$ forms a representation
of $\SU(2)\times\SU(2)$ by translation of the function argument,
\begin{equation}
\label{eq_translationaction}
  (\SU(2)\times\SU(2))\times\Calg(S^3)\to\Calg(S^3),
  ((g_1,g_2),f)\mapsto\bigl(g\mapsto f((g_1,g_2)\ast g)=f(\alignidx{g_1gg_2^{-1}}))\bigr),
\end{equation}
which also gives rise to a representation of
$\SO(4)\cong(\SU(2)\times\SU(2))/\Z_2$.

The decomposition of $\Calg(S^3)$ into irreducible representations of
$\SU(2)\times\SU(2)$ coincides with the Peter--Weyl
decomposition~\eqref{eq_structure_calg} of $\SU(2)$,
\begin{equation}
\label{eq_decomposes3}
  \Calg(S^3)\cong_{\SU(2)\times\SU(2)}\bigoplus_{j\in\frac{1}{2}\N_0}(V_j\otimes V_j^\ast),
\end{equation}
where the direct summand $V_j\otimes V_j^\ast$ for a given $j$ has the
basis
\begin{equation}
  \{\,H^{(j)}_{m_1m_2}\colon\quad 1\leq m_1,m_2\leq 2j+1\,\}.
\end{equation}
In our notation, the spherical functions $H^{(j)}_{mn}$ therefore
transform as the coefficients of a vector in $V_j\otimes
V_j^\ast\cong\Hom_{\SU(2)}(V_j,V_j)$. 
\end{corollary}

Furthermore, the Haar measure of $\SU(2)$ provides an integral over
$S^3$ with the desired translation invariance under the
$\SO(4)$-action.

\begin{proposition}
\label{prop_integrals}
Let $f\in\Calg(S^3)$ and $(g_1,g_2)\in\SU(2)\times\SU(2)$. Then the
integral over $S^3$, obtained from the Haar measure of $\SU(2)$ under
the identification $S^3\cong\SU(2)$, satisfies
\begin{equation}
  \int_{S^3}f(g)\,dg=\int_{S^3}f((g_1,g_2)\ast g)\,dg.
\end{equation}
There is a scalar product on $\Calg(S^3)$,
\begin{equation}
  \left<f_1,f_2\right>:=\int_{S^3}\overline{f_1(g)}f_2(g)\,dg,
\end{equation}
with respect to which the spherical functions are orthogonal,
\begin{equation}
\label{eq_s3ortho}
  \bigl<H^{(j)}_{mn},H^{(k)}_{pq}\bigr>=\frac{1}{2j+1}\delta_{jk}\delta_{mp}\delta_{nq},
\end{equation}
where $j,k\in\frac{1}{2}\N_0$ and $1\leq m,n\leq2j+1$, $1\leq
p,q\leq2k+1$. 
\end{proposition}

We also construct the algebras of functions on $\SU(2)\times\SU(2)$
and $\SO(4)$. We denote the irreducible representations of
$\SU(2)\times\SU(2)$ by pairs $(j_1,j_2)$ of half-integers
$j_1,j_2\in\frac{1}{2}\N_0$. They are of the form
$V_{(j_1,j_2)}=V_{j_1}\otimes V_{j_2}$. A basis for the algebra
$\Calg(\SU(2)\times\SU(2))$ is given by the functions
\begin{equation}
  t^{(j_1,j_2)}_{m_1m_2,n_1n_2}\colon\SU(2)\times\SU(2)\to\C,
    (g_1,g_2)\mapsto t^{(j_1)}_{m_1n_1}(g_1)\cdot
    t^{(j_2)}_{m_2n_2}(g_2),
\end{equation}
where $j_1,j_2\in\frac{1}{2}\N_0$, $1\leq m_\ell,n_\ell\leq
2j_\ell+1$. It is known that the functions
$t^{(j_1,j_2)}_{m_1m_2,n_1n_2}$ are well defined on
$\SO(4)\cong(\SU(2)\times\SU(2))/\Z_2$ if and only if $j_1+j_2\in\N_0$,
\ie\ if in $V_{(j_1,j_2)}=V_{j_1}\otimes V_{j_2}$ either both factors
are even- or both are odd-dimensional. Therefore
$t^{(j_1,j_2)}_{m_1m_2,n_1n_2}\in\Calg(\SO(4))$ if and only if
$j_1+j_2\in\N_0$. 

Usually, the spherical functions of $S^3$ are constructed as the
functions on $\SO(4)/\SO(3)$, \ie\ as the functions on $\SO(4)$ that
are constant on cosets of $\SO(3)$~\cite{ViKl93}. There exist
representation functions $t^{(j_1,j_2)}_{mn,pq}\in\Calg(\SO(4))$ that
are constant on these cosets only if $j_1=j_2$, \ie\ in the class-1
representations of $\SO(4)$ with respect to $\SO(3)$. These are the
\emph{balanced} representations $V_{(j,j)}=V_j\otimes V_j$. 

Our spherical functions~\eqref{eq_spherical}, however, transform as
the coefficients of vectors in $V_{(j,j^\ast)}=V_j\otimes V_j^\ast$.
Recall that for $\SU(2)$, there exist isomorphisms of representations
$V_j\cong V_j^\ast$ (which we have not explicitly constructed) which
can be applied to the right tensor factor in order to relate our
spherical functions to the standard construction.

\subsection{Two-complexes}

\begin{figure}[t]
\begin{center}
\input{pstex/boundary.pstex_t}
\end{center}
\mycaption{fig_boundary}{%
  The maps $\del_\pm$, $\del_j$ and $\epsilon_j$ and the
  conditions~\eqref{eq_boundary}. Here $N(f)=3$, $\epsilon_1f=+1$,
  $\epsilon_2f=+1$ and $\epsilon_3f=(-1)$.}
\end{figure}

In order to define lattice gauge theory and the Barrett--Crane model,
we need essentially an abstract oriented $2$-complex. This is a
collection of faces, edges and vertices together with a boundary
relation specifying which edges are contained in the boundary of a
given face, \etc. For the partition function of lattice gauge theory,
we require a formulation which explicitly describes the cyclic
ordering of the edges in the boundary of a given face. This is
achieved by the following definition of a \emph{$2$-complex with
cyclic structure}.

\begin{definition}
\label{def_complexcyclic}
A \emph{finite $2$-complex with cyclic structure} $(V,E,F)$ consists
of finite sets $V$ (\emph{vertices}), $E$ (\emph{edges}) and $F$
(\emph{faces}) together with maps
\begin{mathletters}
\label{eq_twocomplex}
\begin{alignat}{2}
  \del_+&\colon E\to V,&\qquad&\mbox{(end point of an edge)}\\
  \del_-&\colon E\to V,&\qquad&\mbox{(starting point of an edge)}\\
  N     &\colon F\to\N,&\qquad&\mbox{(number of edges in the boundary
                        of a face)}\\
  \del_j&\colon F\to E,&\qquad&\mbox{(the $j$-th edge in the boundary
                        of a face)}\\
  \epsilon_j&\colon F\to\{-1,+1\},&\qquad&\mbox{(its orientation)}
\end{alignat}%
\end{mathletters}%
so that
\begin{mathletters}
\label{eq_boundary}
\begin{gather}
  \del_{-\epsilon_jf}\del_jf = \del_{\epsilon_{j+1}f}\del_{j+1}f,\qquad
    1\leq j\leq N(f)-1,\\
  \del_{-\epsilon_{N(f)}f}\del_{N(f)}f = \del_{\epsilon_1f}\del_1f,
\end{gather}%
\end{mathletters}%
for all $f\in F$.
\end{definition}

\begin{remark}
\begin{myenumerate}
\item
  The conditions~\eqref{eq_boundary} state that the edges in the
  boundary of a face $f\in F$ are in cyclic ordering from $\del_{N(f)}f$
  to $\del_1f$ where one encounters the edges with a relative
  orientation given by $\epsilon_jf$, see
  Figure~\ref{fig_boundary}. Observe that~\eqref{eq_boundary} contains
  combinatorial information similar to the condition $\del\circ\del=0$
  on the boundary operator $\del$ in Abelian simplicial homology.
\item
  For the existence of the partition functions in the following
  sections, we require that the $2$-complexes have only a finite
  number of vertices, edges and faces.
\end{myenumerate}
\end{remark}

\noindent
In the subsequent calculations, it is convenient to use the following
abbreviations.

\begin{definition}
Let $(V,E,F)$ denote a finite combinatorial $2$-complex with cyclic
structure. For a given edge $e\in E$, the sets
\begin{mathletters}
\begin{eqnarray}
e_+ &:=& \{f\in F\colon\quad e=\del_j f,\quad\epsilon_j f=(+1)\quad
           \mbox{for some $j$,}\quad 1\leq j\leq N(f)\},\\
e_- &:=& \{f\in F\colon\quad e=\del_j f,\quad\epsilon_j f=(-1)\quad
           \mbox{for some $j$,}\quad 1\leq j\leq N(f)\},
\end{eqnarray}%
\end{mathletters}%
contain all faces that have the edge $e$ in their boundary with
positive ($+$) or negative ($-$) orientation, and we write $\delta
e:=e_+\cup e_-$. For a given face $f\in F$, the set
\begin{equation}
  f_0 := \{v\in V\colon\quad v=\del_-\del_jf\quad
           \mbox{for some $j$,}\quad 1\leq j\leq N(f)\},
\end{equation}
denotes all vertices that belong to the boundary of the face
$f$. Finally, the sets 
\begin{mathletters}
\begin{eqnarray}
  f_+ &:=& \{e\in E\colon\quad e=\del_jf,\quad\epsilon_jf=(+1)\quad
             \mbox{for some $j$,}\quad 1\leq j\leq N(f)\},\\
  f_- &:=& \{e\in E\colon\quad e=\del_jf,\quad\epsilon_jf=(-1)\quad
             \mbox{for some $j$,}\quad 1\leq j\leq N(f)\},
\end{eqnarray}%
\end{mathletters}%
contain all edges in the boundary of the face $f$ that occur with
positive ($+$) or negative ($-$) orientation, and $\del f:=f_+\cup
f_-$. Finally, for a given face $f\in F$, $v_+(f)$ denotes the edge
$e\in\del f$ for which $v=\del_+e$, and similarly, $v_-(f)$ the edge
$e\in\del f$ for which $v=\del_-e$.
\end{definition}

\begin{remark}
Given any triangulation of a four-manifold, we can construct a
$2$-complex with cyclic structure that is combinatorially dual to the
triangulation. The vertices $V$ are the four-simplices of the
triangulation, the edges $E$ the tetrahedra and the faces $F$ the
triangles. An edge is contained in the boundary of a given face
whenever the corresponding tetrahedron contains the triangle in its
boundary, \etc. A cyclic structure can then be constructed by
induction. In particular, if the triangulation contains only finitely
many four-simplices, we obtain a finite $2$-complex with cyclic
structure.
\end{remark}

%
\section{The Barrett--Crane model}
%
\label{sect_bcmodel}

When we present the definition of the Barrett--Crane
model~\cite{BaCr98,Ba98a}, we wish to emphasize its structural
similarity to lattice gauge theory with pure gauge fields, in
particular to $\SO(4)$ lattice $BF$-theory. Therefore we recall the
two formulations of lattice gauge theory, the original one using group
integrations and the dual formulation~\cite{OePf01,Pf01} as a spin
foam model before we present the definition of the class of
Barrett--Crane models which we consider in this article.

\subsection{Lattice gauge theory}

For more background on lattice gauge theory, we refer the reader to
text books such as~\cite{Ro92,MoMu94}. The partition function of
lattice gauge theory can be formulated on any $2$-complex with cyclic
structure as follows.

\begin{definition}
Let $G$ be a compact Lie group and $(V,E,F)$ be a finite $2$-complex
with cyclic structure. The \emph{partition function} of lattice gauge
theory is defined as
\begin{equation}
\label{eq_partition}
  Z = \Bigl(\prod_{e\in E}\int_G\,dg_e\Bigr)\,
      \prod_{f\in F}w(g_f),\qquad
  g_f:=g_{\del_1f}^{\epsilon_1f}\cdots g_{\del_{N(f)}f}^{\epsilon_{N(f)}f}.
\end{equation}
Here $w\colon G\to\R$ is the local \emph{Boltzmann weight} which is
required to be a positive and $L^2$-integrable class function of $G$
that satisfies $w(g^{-1})=w(g)$.
\end{definition}

\begin{remark}
\begin{myenumerate}
\item
  Our notation in~\eqref{eq_partition} means that there is one
  integral $\int_G\,dg$ for each edge $e\in E$. We use this notation very
  often for integrals and sums in the subsequent calculations.
\item
  The partition function of lattice gauge theory is an integral over
  the group $G$ for each edge of the $2$-complex so that the set of
  configurations is $G^E$. The Boltzmann weight is calculated for each
  face where $g_f$ denotes the ordered product of group elements
  assigned to the edges in the boundary of the face. This is the place
  where the cyclic structure of the $2$-complex enters the definition.
\item
  The value of the partition function is independent of the choice of
  cyclic structure. The starting point for the cyclic numbering of
  edges in the boundary of a face does not matter because the
  Boltzmann weight is given by a class function and thus invariant
  under cyclic permutation of the factors of $g_f$. Reversal of the
  orientation is also a symmetry because it replaces $g_f$ by
  $g_f^{-1}$.
\item
  For lattice Yang--Mills theory, the Boltzmann weight is given by
  $w(g)=\exp(-s(g))$ in terms of the Euclidean action $s\colon
  G\to\R$. For lattice $BF$-theory~\cite{KaSa00}, however, we have
  $w(g)=\delta(g)$. For a generic Lie group $G$ and a generic
  $2$-complex, the partition function is therefore not well defined.
\end{myenumerate}
\end{remark}

\noindent
The Boltzmann weight of lattice gauge theory is invariant under local
gauge transformations.

\begin{lemma}
Let $h\colon V\to G,v\mapsto h_v$ associate a group element to each
vertex. The Boltzmann weight $w(g_f)$ is invariant under the
\emph{local gauge transformations},
\begin{equation}
  G^E\to G^E,g_e\mapsto\alignidx{h_{\del_+e}\cdot g_e\cdot
  h_{\del_-e}^{-1}}.
\end{equation}
\end{lemma}

\begin{proof}
In order to prove this invariance, one has to make use of the
conditions~\eqref{eq_boundary} and of the fact that $w$ is a class
function.
\end{proof}

\begin{figure}[t]
\begin{center}
\input{pstex/transform.pstex_t}
\end{center}
\mycaption{fig_transform}{%
  An edge $e\in E$ in the boundary of three faces, two triangles and
  one quadrilateral. (a) Equation~\eqref{eq_step20} contains one Haar
  intertwiner for each edge. The intertwiners are represented by the
  boxes as in Figure~\ref{fig_diagrams}(d), and the diagram indicates
  how to contract the indices. (b) The contraction of indices in the
  Barrett--Crane model~\eqref{eq_bcmodel}. The full dots represent the
  Barrett--Crane intertwiners~\eqref{eq_bcinter}. We have omitted all
  arrows in both diagrams (a) and (b).}
\end{figure}

There exists a dual formulation of lattice gauge theory in which the
partition function is no longer given by group integrals but rather by
sums over representations and intertwiners~\cite{OePf01,Pf01,Oe02}, so
that it forms a spin foam model. We present a definition of the dual
model which coincides with an intermediate step in~\cite{OePf01,Pf01}
and directly corresponds to the diagrammatical formulation
of~\cite{Oe02}.

\begin{theorem}
\label{thm_dualpart}
Let $G$ be a compact Lie group and $(V,E,F)$ be a finite $2$-complex
with cyclic structure. The partition function of lattice gauge
theory~\eqref{eq_partition} is equal to
\begin{eqnarray}
\label{eq_step20}
  Z &=& \Bigl(\prod_{f\in F}\sum_{\rho_f\in\Irrep}\Bigr)\,
        \Bigl(\prod_{f\in F}\hat w_{\rho_f}\Bigr)\,
        \Bigl(\prod_{f\in F}\prod_{v\in f_0}\sum_{n(f,v)=1}^{\dim V_{\rho_f}}\Bigr)\nn\\
    &&\times\prod_{e\in E}
      T^{(e)}_{\underbrace{n(f,\del_+e)\ldots}_{f\in e_+}
               \underbrace{n(f,\del_+e)\ldots}_{f\in e_-};
               \underbrace{n(f,\del_-e)\ldots}_{f\in e_+}
               \underbrace{n(f,\del_-e)\ldots}_{f\in e_-}},
\end{eqnarray}
where $T^{(e)}$ denotes the Haar intertwiner~\eqref{eq_haarinter} for
the following representations,
\begin{equation}
  T^{(e)}\colon\Bigl(\bigotimes_{f\in e_+}V_{\rho_f}\Bigr)\otimes
               \Bigl(\bigotimes_{f\in e_-}V_{\rho_f}^\ast\Bigr)\to
               \Bigl(\bigotimes_{f\in e_+}V_{\rho_f}\Bigr)\otimes
               \Bigl(\bigotimes_{f\in e_-}V_{\rho_f}^\ast\Bigr).
\end{equation}
We require here that the ordering of tensor factors parameterized by
$f\in e_\pm$ agrees with the ordering of indices
in~\eqref{eq_step20}. The coefficients $\hat w_{\rho}$ are the
coefficients of the character expansion of the Boltzmann weight,
\begin{equation}
  \hat w_\rho=\int_G\overline{\chi^{(\rho)}(g)}w(g)\,dg.
\end{equation}
\end{theorem}

\begin{remark}
\begin{myenumerate}
\item
  The dual form of the partition function is a sum over all colourings
  of faces $f\in F$ with irreducible representations $\rho\in\Irrep$
  of $G$. There is a Haar intertwiner $T^{(e)}$ assigned to each edge
  $e\in E$, and under the sum we have a weight $\hat w_\rho$ for each
  face. The sums over the indices $n(f,v)$ for each face $f\in F$ and
  each vertex $v\in f_0$ in the boundary of the face, contract the
  indices of the Haar intertwiners. This is illustrated in
  Figure~\ref{fig_transform}(a). 
\item
  For lattice $BF$-theory, we have $\hat w_\rho=\dim V_\rho$, \ie\ the
  sums do not converge. In particular, for $G=\SU(2)$ and a
  $2$-complex that is dual to a given triangulation of a
  four-manifold, we obtain the Ooguri model~\cite{Oo92} whose $\SO(4)$
  version we will consider later.
\item
  For lattice Yang--Mills theory, the transformation
  relating~\eqref{eq_partition} to~\eqref{eq_step20} is a strong-weak
  or high temperature-low temperature duality
  transformation~\cite{OePf01,PfOe02}. This is a consequence of the
  fact that the character expansion of the Boltzmann weight has many
  similarities to Fourier decomposition and maps a Boltzmann weight
  with a wide peak to a function with a narrow peak and conversely.
\end{myenumerate}
\end{remark}

\subsection{The Barrett--Crane intertwiner}

The Barrett--Crane model has close similarities to the spin foam model
dual to $\SO(4)$ lattice $BF$-theory. We obtain the Barrett--Crane
model starting from the partition function~\eqref{eq_step20} if we
restrict the sum over irreducible representations to the balanced
representations of $\SO(4)$. Furthermore, we have to replace the Haar
intertwiner of~\eqref{eq_step20} by a pair of Barrett--Crane
intertwiners.

Following Barrett, Freidel and Krasnov~\cite{Ba98b,FrKr00}, the
Barrett--Crane intertwiner can be written as an integral over $S^3$.

\begin{definition}
\label{def_bcinter}
Let $V_{(j_\ell,j_\ell^\ast)}$ and $V_{(k_\ell,k_\ell^\ast)}$,
$j_\ell,k_\ell\in\frac{1}{2}\N_0$, denote balanced irreducible
representations of $\SO(4)$ whose right tensor factor has been
dualized. The \emph{Barrett--Crane intertwiner}
\begin{mathletters}
\label{eq_bcinter}
\begin{equation}
  I\colon\bigotimes_{\ell=1}^r \alignidx{V_{(j_\ell,j_\ell^\ast)}}\to 
         \bigotimes_{\ell=1}^s \alignidx{V_{(k_\ell,k_\ell^\ast)}},
\end{equation}%
is defined in terms of the matrix elements,
\begin{equation}
\label{eq_bcintermatrix}
  I_{(m_1n_1)\ldots;(p_1q_1)\ldots} := \int_{S^3}
    \overline{H^{(j_1)}_{p_1q_1}(x)\cdots H^{(j_r)}_{p_rq_r}(x)}
              H^{(k_1)}_{m_1n_1}(x)\cdots H^{(k_s)}_{m_sn_s}(x)\,dx,
\end{equation}%
\end{mathletters}%
of the intertwiner
\begin{equation}
  \C\to\Bigl(\bigotimes_{\ell=1}^r\alignidx{V_{(j_\ell,j_\ell^\ast)}^\ast}\Bigr)
       \otimes\Bigl(\bigotimes_{\ell=1}^s\alignidx{V_{(k_\ell,k_\ell^\ast)}}\Bigr).
\end{equation}
Here the identification
$\Hom_{\SO(4)}(U,W)\cong\Hom_{\SO(4)}(\C,U^\ast\otimes W)$ was used as
well as the fact that $\overline{H^{(j)}_{pq}}=H^{(j^\ast)}_{pq}$ and
$V_{(j,j^\ast)}^\ast\cong V_j^\ast\otimes{V_j^\ast}^\ast$. The index
pairs $(p_\ell q_\ell)$ therefore correspond to the tensor factors
$\alignidx{V_{(j_\ell,j_\ell^\ast)}}$ of the domain of $I$ while the
$(m_\ell n_\ell)$ belong to the image
$\alignidx{V_{(k_\ell,k_\ell^\ast)}}$.
\end{definition}

\begin{lemma}
The map $I$ defined by~\eqref{eq_bcinter} is a morphism of
representations of $\SO(4)$.
\end{lemma}

\begin{proof}
The action of $\SO(4)$ is given by the action of $\SU(2)\times\SU(2)$
in~\eqref{eq_translationaction}. If we write $\SU(2)\cong S^3$ for the
range of the integral, 
\begin{equation}
  I_{(m_1n_1)\ldots;(p_1q_1)\ldots} := \int_{\SU(2)}
    t^{(j_1^\ast)}_{p_1q_1}(g)\cdots t^{(j_r^\ast)}_{p_rq_r}(g)
    t^{(k_1)}_{m_1n_1}(g)\cdots t^{(k_s)}_{m_sn_s}(g)\,dg,
\end{equation}
this action corresponds to $\SU(2)$ left-right
translation of all function arguments which is an invariance of the
Haar measure of $\SU(2)$.
\end{proof}

\subsection{Definition of the model}

The Barrett--Crane model can now be defined by an expression similar
to the dual of lattice gauge theory (Theorem~\ref{thm_dualpart})
in which the Haar intertwiner is replaced by a pair of Barrett--Crane
intertwiners, one associated to each vertex of the given
edge. These two Barrett--Crane intertwiners are `independent' which
corresponds to the first version of the model proposed
in~\cite{BaCr98}. 

\begin{definition}
Let $(V,E,F)$ denote a finite $2$-complex with cyclic structure, let
$\hat w_j\in\C$, $j\in\frac{1}{2}\N_0$, be a face amplitude and
\begin{equation}
  \sym{A}^{(e)}(\underbrace{j_f,\ldots}_{f\in\delta e})\in\C,
\end{equation}
$j_f\in\frac{1}{2}N_0$, an edge amplitude. The Barrett--Crane model is
defined by the partition function,
\begin{eqnarray}
\label{eq_bcmodel}
  Z &=& \Bigl(\prod_{f\in F}\sum_{j_f\in\frac{1}{2}\N_0}\Bigr)\,
        \Bigl(\prod_{f\in F}\hat w_{j_f}\Bigr)\,
        \Bigl(\prod_{e\in E}\sym{A}^{(e)}(\underbrace{j_f,\ldots}_{f\in\delta e})\Bigr)\,
        \Bigl(\prod_{f\in F}\prod_{v\in f_0}\,\sum_{n(f,v)=1}^{2j_f+1}\,\sum_{m(f,v)=1}^{2j_f+1}\Bigr)\\
  &&\times\prod_{e\in E} I^{(+,e)}_{
     \underbrace{\scriptstyle (n(f,\del_+e)m(f,\del_+e))\ldots}_{f\in e_+};
     \underbrace{\scriptstyle (n(f,\del_+e)m(f,\del_+e))\ldots}_{f\in e_-}}\cdot
                         I^{(-,e)}_{
     \underbrace{\scriptstyle (n(f,\del_-e)m(f,\del_-e))\ldots}_{f\in e_-};
     \underbrace{\scriptstyle (n(f,\del_-e)m(f,\del_-e))\ldots}_{f\in e_+}}.\nn
\end{eqnarray}
Here $I^{(+,e)}_{\ldots}$ and $I^{(-,e)}_{\ldots}$, $e\in E$, denote
the coefficients of the Barrett--Crane intertwiners~\eqref{eq_bcinter},
\begin{mathletters}%
\label{eq_bcintermap}%
\begin{eqnarray}%
  I^{(+,e)}&\colon&\bigotimes_{f\in e_-}\alignidx{V_{(j_f,j_f^\ast)}}\to
                   \bigotimes_{f\in e_+}\alignidx{V_{(j_f,j_f^\ast)}},\\
  I^{(-,e)}&\colon&\bigotimes_{f\in e_+}\alignidx{V_{(j_f,j_f^\ast)}}\to
                   \bigotimes_{f\in e_-}\alignidx{V_{(j_f,j_f^\ast)}},
\end{eqnarray}%
\end{mathletters}%
where $V_{(j,j^\ast)}=\alignidx{V_j\otimes V_j^\ast}$ are balanced
irreducible representations of $\SO(4)$ whose right factors have been
dualized.
\end{definition}

\begin{figure}[t]
\begin{center}
\input{pstex/bcinter.pstex_t}
\end{center}
\mycaption{fig_bcinter}{%
  (a) Replacing the $\SO(4)$ Haar intertwiner by a pair of
  Barrett--Crane intertwiners~\eqref{eq_bcinter}. The lines labelled
  $j$ represent balanced irreducible representations $V_{(j,j^\ast)}$ of
  $\SO(4)$. (b) Commutativity of the algebra of spherical
  functions. (c) Relation of the conjugate spherical function with the
  dual representation.}
\end{figure}

\begin{figure}[t]
\begin{center}
\input{pstex/trivalent.pstex_t}
\end{center}
\mycaption{fig_trivalent}{%
  (a) Trivalent vertices represent the recoupling coefficients of the
  spherical functions. (b) Commutativity of the algebra. (c) Linearity
  of the integral. (d) The orthogonality relation for spherical
  functions.}
\end{figure}

\begin{remark}
\begin{myenumerate}
\item
  The partition function~\eqref{eq_bcmodel} can be described in words
  as follows. The faces $f\in F$ are coloured in all possible ways
  with representations $V_{(j,j^\ast)}=\alignidx{V_j\otimes
  V_j^\ast}$, $j\in\frac{1}{2}\N_0$, of $\SO(4)$. There is an
  amplitude $\hat w_{j_f}$ for each face and an amplitude
  $\sym{A}^{(e)}(j_f,\ldots)$ for each edge which can depend on the
  representations at all faces $f\in\delta e$. Each edge $e\in E$ is
  labelled with a pair of Barrett--Crane intertwiners $I^{(+,e)}$ and
  $I^{(-,e)}$, each one associated to one vertex, and the vector
  indices (here pairs $(m,n)$ for $V_j\otimes V_j^\ast$) are
  contracted precisely as in the spin foam model dual to lattice gauge
  theory (Theorem~\ref{thm_dualpart} and
  Figure~\ref{fig_transform}(a)). Figure~\ref{fig_bcinter}(a)
  visualizes the replacement of the $\SO(4)$ Haar intertwiner $T$ by
  the pair of Barrett--Crane intertwiners $I$. The contraction of the
  indices in the Barrett--Crane model is illustrated in
  Figure~\ref{fig_transform}(b). In this figure, we denote the
  Barrett--Crane intertwiners by the full dots, a simplification which
  we justify below. 
\item
  The face amplitude is usually chosen to be $\hat w_j=\dim
  V_{(j,j^\ast)}={(2j+1)}^2$ while there are various edge amplitudes
  proposed in the literature~\cite{Ba98a,DPFr00,OrWi01,PeRo01}. We
  leave the amplitudes unspecified here and discuss the question of
  the convergence of the partition function later in
  Section~\ref{sect_bcregular}. 
\item
  For all balanced irreducible representations $V_j\otimes V_j$ of
  $\SU(2)\times\SU(2)$, we have identified the tensor factor
  corresponding to the right $\SU(2)$ by its dual. Our model still
  agrees with the original proposal~\cite{BaCr98} since the
  corresponding isomorphism appears in each representation of
  $\SO(4)$, \ie\ in each line in Figure~\ref{fig_transform}(b),
  together with its inverse in such a way that they cancel. We have,
  however, obtained a considerable technical advantage with this
  construction as we can now use the spherical functions of
  Section~\ref{sect_s3} which agree with the representation functions
  of $\SU(2)$.
\end{myenumerate}
\end{remark}

\subsection{Properties of the Barrett--Crane intertwiner}

\begin{figure}[t]
\begin{center}
\input{pstex/bcsu2.pstex_t}
\end{center}
\mycaption{fig_bcsu2}{%
  (a) Expressing representations $V_{(j,j^\ast)}=\alignidx{V_j\otimes
  V_j^\ast}$ of $\SO(4)$ as a tensor product of representations of
  $\SU(2)$. The tensor factor corresponding to the right $\SU(2)$ is
  drawn as a dotted line. (b) The Barrett--Crane intertwiner as an
  $\SU(2)$ integral~\eqref{eq_bcsu2}.} 
\end{figure}

In this section, we comment on a number of properties of the
Barrett--Crane intertwiner~\eqref{eq_bcinter} and translate some
results of~\cite{Ba98b} into our language. This intertwiner is
visualized by the box diagrams labelled $I$ in
Figure~\ref{fig_bcinter}(a). Each line carries a label
$j\in\frac{1}{2}\N_0$ and symbolizes the representation
$\alignidx{V_j\otimes V_j^\ast}$ of $\SU(2)\times\SU(2)$ if the arrow
points down and the dual representation ${(\alignidx{V_j\otimes
V_j^\ast})}^\ast$ if the arrow points up.

Commutativity of the algebra $\Calg(S^3)$ gives rise to rules such as
the one depicted in Figure~\ref{fig_bcinter}(b). If we use
$\overline{H^{(j)}_{mn}}=H^{(j^\ast)}_{mn}$ in the integrand, we
obtain rules as in Figure~\ref{fig_bcinter}(c). The lines can
therefore be freely moved around the diagram, and what matters is only
whether the arrows point towards or away from the box. Therefore, in
Figure~\ref{fig_transform}(b), it is sufficient to draw full dots with
incoming or outgoing lines for the Barrett--Crane intertwiners. Note
that each arrow can be reversed if at the same time the representation
label $j$ is replaced by its dual $j^\ast$.

We can make use of the fact that $\Calg(S^3)$ forms an algebra for
which the spherical functions provide a basis
(Corollary~\ref{corr_decomposes3}), therefore, 
\begin{mathletters}
\begin{eqnarray}
\label{eq_trivalent}
  H^{(j_1)}_{m_1n_1}\cdot H^{(j_2)}_{m_2n_2}
  &=& \sum_{k\in\frac{1}{2}\N_0}\sum_{p,q=1}^{2k+1}
        \sqrt{\frac{2k+1}{(2j_1+1)(2j_2+1)}}\cdot 
           c^{(j_1j_2;k)}_{m_1n_1,m_2n_2;pq}\cdot H^{(k)}_{pq},\\
  \overline{H^{(j_1)}_{m_1n_1}}\cdot\overline{H^{(j_2)}_{m_2n_2}}
  &=& \sum_{k\in\frac{1}{2}\N_0}\sum_{p,q=1}^{2k+1}
        \sqrt{\frac{2k+1}{(2j_1+1)(2j_2+1)}}\cdot
           \overline{c^{(j_1j_2;k)}_{m_1n_1,m_2n_2;pq}}\cdot
           \overline{H^{(k)}_{pq}},
\end{eqnarray}%
\end{mathletters}%
for $j_1,j_2\in\frac{1}{2}\N_0$ and $1\leq m_\ell,n_\ell\leq
2j_\ell+1$, with uniquely determined coefficients
$c^{(j_1j_2;k)}_{m_1n_1,m_2n_2;pq}\in\C$. This is illustrated in
Figure~\ref{fig_trivalent}(a). Rules such as the one depicted in
diagram~(b) follow from the commutativity of the algebra. Linearity of
the integral implies rules as in diagram~(c), and the orthogonality of
spherical functions~\eqref{eq_s3ortho} finally leads to diagram~(d).

These rules can be used in order to decompose the Barrett--Crane
intertwiner into a tree diagram involving only three-valent
vertices~\cite{Ba98b}. For each line connecting two vertices, the
corresponding indices are contracted, and we sum over all
representations of the form $\alignidx{V_j\otimes V_j^\ast}$ for each
internal edge. Obviously, all possible decompositions into trees
evaluate to the same intertwiner~\eqref{eq_bcinter} \cite{Ba98b}. This is the
property that was originally required in~\cite{BaCr98,Re99}. We
emphasize that we here use the intertwiner in the form involving the
spherical functions $H^{(j)}_{mn}$ whose orthogonality
relation~\eqref{eq_s3ortho} involves a denominator of $2j+1$. The
normalized basis vectors of $\Calg(S^3)$ are therefore
$\sqrt{2j+1}\cdot H^{(j)}_{mn}$.

\begin{figure}[t]
\begin{center}
\input{pstex/normalize.pstex_t}
\end{center}
\mycaption{fig_normalize}{%
  Calculation in order to determine the normalization of the Barrett--Crane
  intertwiner~\eqref{eq_bcinter}. The Barrett--Crane intertwiner is
  denoted by the boxes labelled $I$ while the boxes $T$ represent the
  $\SU(2)$ Haar intertwiner. In the first step, we have
  used~\eqref{eq_bcsu2} (Figure~\ref{fig_bcsu2}(b)), and in the second
  step the projector property~\eqref{eq_interproj} of the Haar
  intertwiner.}
\end{figure}

Furthermore, we can employ the identification of the $S^3$-integral of
spherical functions with the $\SU(2)$-integral of representation
functions (Proposition~\ref{prop_integrals}) and express the
Barrett--Crane intertwiner as an $\SU(2)$ integral,
\begin{mathletters}
\begin{eqnarray}
\label{eq_bcsu2}
  &&\int_{S^3}\overline{H^{(j_1)}_{p_1q_1}(x)\cdots H^{(j_r)}_{p_rq_r}(x)}\cdot
                      H^{(k_1)}_{m_1n_1}(x)\cdots H^{(k_s)}_{m_sn_s}(x)\,dx\nn\\
  &=& \int_{\SU(2)}t^{(j_1^\ast)}_{p_1q_1}(g)\cdots t^{(j_r^\ast)}_{p_rq_r}(g)\cdot
                 t^{(k_1)}_{m_1n_1}(g)\cdots t^{(k_s)}_{m_sn_s}(g)\,dg,
\end{eqnarray}%
so that we find
\label{eq_bcsu2b}
\begin{equation}
  I_{(m_1n_1)\ldots;(p_1q_1)\ldots} = T_{p_1\ldots m_1\ldots;q_1\ldots n_1\ldots},
\end{equation}%
\end{mathletters}%
where $T$ is the $\SU(2)$ Haar intertwiner,
\begin{equation}
  T\colon\Bigl(\bigotimes_{\ell=1}^sV_{k_\ell}\Bigr)\otimes
         \Bigl(\bigotimes_{\ell=1}^rV_{j_\ell}^\ast\Bigr)\to
         \Bigl(\bigotimes_{\ell=1}^sV_{k_\ell}\Bigr)\otimes
         \Bigl(\bigotimes_{\ell=1}^rV_{j_\ell}^\ast\Bigr).
\end{equation}

This relation is shown in Figure~\ref{fig_bcsu2}. In diagram~(a), the
representations $V_{(j,j^\ast)}\cong\alignidx{V_j\otimes V_j^\ast}$ of
$\SO(4)$ are split into their tensor factors as representations of
$\SU(2)\times\SU(2)$ where we have dualized all right
factors. Diagram~(b) visualizes~\eqref{eq_bcsu2b}.

Finally, we show that the Barrett--Crane intertwiner of
Definition~\ref{def_bcinter} agrees with the unnormalized intertwiner
in the terminology of~\cite{DPFr00}. Therefore we consider the
intertwiners $I^{(1)}\colon\C\to
V_{(j_1,j_1^\ast)}\otimes\cdots\otimes V_{(j_4,j_4^\ast)}$ and
$I^{(2)}\colon V_{(j_1,j_1^\ast)}\otimes\cdots\otimes
V_{(j_4,j_4^\ast)}\to\C$ for a four-fold tensor product and compute
\begin{eqnarray}
  &&\sum_{m_1,n_1=1}^{2j_1+1}\cdots\sum_{m_4,n_4=1}^{2j_4+1} 
    I^{(1)}_{(m_1n_1)\cdots(m_4n_4)}\cdot I^{(2)}_{(m_1n_1)\ldots(m_4n_4)}\nn\\
  &=&\int_{\SU(2)}\chi^{(j_1)}(g)\chi^{(j_2)}(g)\chi^{(j_3)}(g)\chi^{(j_4)}(g)\,dg\nn\\
  &=&\dim\Hom_{\SU(2)}(V_{j_1}\otimes V_{j_2},V_{j_3}\otimes V_{j_4}).
\end{eqnarray}
This calculation is illustrated in Figure~\ref{fig_normalize}. A
comparison shows that our Barrett--Crane intertwiner agrees with the
unnormalized one of~\cite{DPFr00}. If we wish to use a different
normalization and multiply our intertwiner by a factor, we can always
absorb this factor into the edge amplitude $\sym{A}^{(e)}$. We
emphasize that the Barrett--Crane intertwiner for given
representations is unique only up to normalization~\cite{Re99}, but
that the relative weight of different representations is not
determined. We demonstrate in Section~\ref{sect_bcregular} that the
physical properties of the model are affected by the choice of
normalization.

%
\section{Dual variables for the Barrett--Crane model}
%
\label{sect_transform}

\subsection{The duality transformation}

We can employ the integral presentation of the Barrett--Crane
intertwiner~\eqref{eq_bcinter} in order to introduce $S^3$-variables
into the Barrett--Crane model. It is possible to derive a
transformation similar to the reversed direction of the duality
transformation for lattice gauge theory in order to promote the new
$S^3$-values to the configuration variables of the partition function.

First we make use of the integral presentation of the Barrett--Crane
intertwiner,
\begin{eqnarray}
\label{eq_bcintersu2}
  &&I^{(+,e)}_{\underbrace{\scriptstyle n(f,\del_+e)m(f,\del_+e)\ldots}_{f\in e_+};
               \underbrace{\scriptstyle n(f,\del_+e)m(f,\del_+e)\ldots}_{f\in e_-}}\cdot
    I^{(-,e)}_{\underbrace{\scriptstyle n(f,\del_-e)m(f,\del_-e)\ldots}_{f\in e_-};
               \underbrace{\scriptstyle n(f,\del_-e)m(f,\del_-e)\ldots}_{f\in e_+}}\nn\\
  &=& \int_{\SU(2)}
               \underbrace{t^{(j_f)}_{n(f,\del_+e)m(f,\del_+e)}(g)\cdots}_{f\in e_+}
               \underbrace{t^{(j_f^\ast)}_{n(f,\del_+e)m(f,\del_+e)}(g)\cdots}_{f\in e_-}\,dg\nn\\
  &&\times\int_{\SU(2)}
               \underbrace{t^{(j_f)}_{n(f,\del_-e)m(f,\del_-e)}(g)\cdots}_{f\in e_-}
               \underbrace{t^{(j_f^\ast)}_{n(f,\del_-e)m(f,\del_-e)}(g)\cdots}_{f\in e_+}\,dg.
\end{eqnarray}
We insert this expression into the partition
function~\eqref{eq_bcmodel} and move all integrals to the left,
\begin{eqnarray}
\label{eq_step12}
  Z &=& \Bigl(\prod_{e\in E}\int_{\SU(2)}\,dg^{(+)}_e\int_{\SU(2)}\,dg^{(-)}_e\Bigr)\,
        \Bigl(\prod_{f\in F}\sum_{j_f\in\frac{1}{2}\N_0}\Bigr)\,
        \Bigl(\prod_{f\in F}\hat w_{j_f}\Bigr)\,
        \Bigl(\prod_{e\in E}\sym{A}^{(e)}(\underbrace{j_f,\ldots}_{f\in\delta e})\Bigr)\nn\\
    &&\times \Bigl(\prod_{f\in F}\prod_{v\in f_0}\sum_{n(f,v)=1}^{2j_f+1}\sum_{m(f,v)=1}^{2j_f+1}\Bigr)\nn\\
    &&\times\prod_{e\in E}\Biggl[
        \Bigl(\prod_{f\in e_-}t^{(j_f)}_{m(f,\del_+e)n(f,\del_+e)}({g_e^{(+)}}^{-1})\cdot
                              t^{(j_f)}_{n(f,\del_-e)m(f,\del_-e)}(g_e^{(-)})\Bigr)\nn\\
    &&\qquad\Bigl(\prod_{f\in e_+}t^{(j_f)}_{n(f,\del_+e)m(f,\del_+e)}(g_e^{(+)})\cdot
                                  t^{(j_f)}_{m(f,\del_-e)n(f,\del_-e)}({g_e^{(-)}}^{-1})\Bigr)\Biggr].
\end{eqnarray}
Here we have also employed the antipode~\eqref{eq_antipode} in order
to remove dual representations in favour of inverse group elements.

Keeping the similarity with lattice gauge theory in mind, we try to
factor the integrand into one contribution for each face. Therefore,
we sort the products in the above expression by face rather than by
edge. This step is almost entirely hidden in our notation,
\begin{eqnarray}
  Z &=& \Bigl(\prod_{e\in E}\int_{\SU(2)}\,dg^{(+)}_e\int_{\SU(2)}\,dg^{(-)}_e\Bigr)\,
        \Bigl(\prod_{f\in F}\sum_{j_f\in\frac{1}{2}\N_0}\Bigr)\,
        \Bigl(\prod_{f\in F}\hat w_{j_f}\Bigr)\,
        \Bigl(\prod_{e\in E}\sym{A}^{(e)}(\underbrace{j_f,\ldots}_{f\in\delta e})\Bigr)\nn\\
    &&\times \Bigl(\prod_{f\in F}\prod_{v\in f_0}\sum_{n(f,v)=1}^{2j_f+1}\sum_{m(f,v)=1}^{2j_f+1}\Bigr)\nn\\
    &&\times\prod_{f\in F}\Biggl[
        \Bigl(\prod_{e\in f_-}t^{(j_f)}_{m(f,\del_+e)n(f,\del_+e)}({g_e^{(+)}}^{-1})\cdot
                              t^{(j_f)}_{n(f,\del_-e)m(f,\del_-e)}(g_e^{(-)})\Bigr)\nn\\
    &&\qquad\Bigl(\prod_{e\in f_+}t^{(j_f)}_{n(f,\del_+e)m(f,\del_+e)}(g_e^{(+)})\cdot
                                  t^{(j_f)}_{m(f,\del_-e)n(f,\del_-e)}({g_e^{(-)}}^{-1})\Bigr)\Biggr].
\end{eqnarray}
The structure of the integrand becomes clearer if we re-introduce the
cyclic ordering of edges in the boundary of each face,
\begin{eqnarray}
\label{eq_step10}
  Z &=& \Bigl(\prod_{e\in E}\int_{\SU(2)}\,dg_e^{(+)}\int_{\SU(2)}\,dg_e^{(-)}\Bigr)\,
        \Bigl(\prod_{f\in F}\sum_{j_f\in\frac{1}{2}\N_0}\Bigr)\,
        \Bigl(\prod_{e\in E}\sym{A}^{(e)}(\underbrace{j_f,\ldots}_{f\in\delta e})\Bigr)\nn\\
    &&\times \prod_{f\in F}\Biggl[\hat w_{j_f}\,\prod_{v\in f_0}\sum_{n(v)=1}^{2j_f+1}\sum_{m(v)=1}^{2j_f+1}\Biggr.\nn\\
  &&\quad\Biggl.\prod_{k=1}^{N(f)}
        t_{m(\del_{-\epsilon_k}\del_k f)n(\del_{-\epsilon_k}\del_k f)}({g_{\del_k f}^{(\epsilon_k f)}}^{-1})\cdot
        t_{n(\del_{\epsilon_k}\del_k f)m(\del_{\epsilon_k}\del_k f)}(g_{\del_k f}^{(\epsilon_k f)})\Biggr].
\end{eqnarray}
Recall that the cyclic structure of the edges around a face is
described by the conditions~\eqref{eq_boundary}. We notice that in the
last product in~\eqref{eq_step10}, both indices of the second
$t^{(j)}_{nm}$ of each factor are contracted with the indices of the
first $t^{(j)}_{nm}$ of the following factor and so on.

We therefore shift the factors of the product by a single
$t^{(j)}_{nm}$. Now the indices $n(v)$ and $m(v)$ occur only in a
single factor, 
\begin{eqnarray}
\label{eq_step11}
  Z &=& \Bigl(\prod_{e\in E}\int_{\SU(2)}\,dg_e^{(+)}\int_{\SU(2)}\,dg_e^{(-)}\Bigr)\,
      \Bigl(\prod_{f\in F}\sum_{j_f\in\frac{1}{2}\N_0}\Bigr)\,
      \Bigl(\prod_{e\in E}\sym{A}^{(e)}(\underbrace{j_f,\ldots}_{f\in\delta e})\Bigr)\nn\\
  &&\times\prod_{f\in F}\Biggl[\hat w_{j_f}\,\prod_{v\in f_0}
        \sum_{n=1}^{2j_f+1}\sum_{m=1}^{2j_f+1}
          t^{(j_f)}_{nm}(g_{v_+(f)}^{(+)})\,t^{(j_f)}_{mn}({g_{v_-(f)}^{(-)}}^{-1})\Biggr],
\end{eqnarray}
where $v_+(f)$ denotes the edge in the boundary of the face $f\in F$
for which $v=\del_+e$, \etc. The last expression involves
just a group character. We arrive at the following result.

\begin{theorem}[Dual variables]
\label{thm_bcdual}
Let $(V,E,F)$ be a finite $2$-complex with cyclic structure, let $\hat
w_j\in\C$, $j\in\frac{1}{2}\N_0$, be a face amplitude and
$\sym{A}^{(e)}(j_f,\ldots)$ denote an edge amplitude,
$j_f\in\frac{1}{2}\N_0$. The partition function of the
Barrett--Crane model~\eqref{eq_bcmodel} is equal to the following
expression,
\begin{eqnarray}
\label{eq_bcintermediate}
Z &=& \Bigl(\prod_{e\in E}\int_{\SU(2)}\,dg_e^{(+)}\int_{\SU(2)}\,dg_e^{(-)}\Bigr)\,
    \Bigl(\prod_{f\in F}\sum_{j_f\in\frac{1}{2}\N_0}\Bigr)\,
    \Bigl(\prod_{e\in E}\sym{A}^{(e)}(\underbrace{j_f,\ldots}_{f\in\delta e})\Bigr)\nn\\
  &&\times \prod_{f\in F}\Bigl[\hat w_{j_f}
      \prod_{v\in f_0}\chi^{(j_f)}(g_{v_+(f)}^{(+)}\cdot{g_{v_-(f)}^{(-)}}^{-1})\Bigr].
\end{eqnarray}%
If the edge amplitude factors into one contribution for each face
attached to the edge,
\begin{equation}
\label{eq_edgefactor}
  \sym{A}^{(e)}(\underbrace{j_f,\ldots}_{f\in\delta e})=\prod_{f\in\delta e}\hat v_{j_f},
\end{equation}
$\hat v_{j_f}\in\C$, this can be written as
\begin{mathletters}
\begin{equation}
\label{eq_bcdual}
Z = \Bigl(\prod_{e\in E}\int_{\SU(2)}\,dg_e^{(+)}\int_{\SU(2)}\,dg_e^{(-)}\Bigr)\,
    \prod_{f\in F}f(g_{\del_1f}^{(+)},g_{\del_1f}^{(-)},
      g_{\del_2f}^{(+)},g_{\del_2f}^{(-)},\ldots,g_{\del_{N(f)}f}^{(+)},g_{\del_{N(f)}f}^{(-)}),
\end{equation}%
where
\begin{equation}
\label{eq_bcweight}
  f(g_1^{(+)},g_1^{(-)},\ldots)=\sum_{j\in\frac{1}{2}\N_0}\alignidx{\hat w_j\,\hat v_j^{|\del f|}}
    \chi^{(j)}(\alignidx{g_2^{(+)}{g_1^{(-)}}^{-1}})
    \chi^{(j)}(\alignidx{g_3^{(+)}{g_2^{(-)}}^{-1}})\cdots
    \chi^{(j)}(\alignidx{g_1^{(+)}{g_k^{(-)}}^{-1}}).
\end{equation}%
\end{mathletters}%
\end{theorem}

\begin{remark}
\begin{myenumerate}
\item
  The partition function~\eqref{eq_bcintermediate} describes a system in
  Statistical Mechanics whose variables are pairs of $\SU(2)$ values
  $(g_e^{(+)},g_e^{(-)})$ attached to each edge $e\in E$. For a generic
  edge amplitude $\sym{A}^{(e)}$, the Boltzmann weight under the
  partition sum forms a non-local expression.
\item
  If the edge amplitude factorizes into one contribution per face, the
  Boltzmann weight factorizes into a product over all
  faces~\eqref{eq_bcdual}. The factor
  $f(g_1^{(+)},g_1^{(-)},\ldots,g_k^{(+)},g_k^{(-)})$ for each face
  $f\in F$ depends on the group elements at all edges in the boundary
  of $f$, but only through the characters
  $\chi^{(j)}(\alignidx{g_{\ell+1}^{(+)}\cdot{g_\ell^{(-)}}^{-1}})$
  for each pair of group elements $(g_{\ell+1}^{(+)},g_\ell^{(-)})$
  that are associated to neighbouring edges. Although this interaction
  is considerably more complicated than just a nearest neighbour
  interaction, it is still local in the sense that there is one
  contribution for each face, and the interaction term for a given face
  relies only on data associated to the nearby edges.
\item
  We discuss the properties of the Boltzmann weight
  $f(g_1^{(+)},g_1^{(-)},\ldots)$ and the possible choices for the
  amplitudes $w_j$ and $\sym{A}^{(e)}$ in
  Section~\ref{sect_bcregular}.
\item
  Equation~\eqref{eq_bcintermediate} is an intermediate step in the
  transformation in which both types of variables are present, the old
  representation numbers and the new group variables.
\end{myenumerate}
\end{remark}

\subsection{Geometric interpretation}

\begin{figure}[t!]
\begin{center}
\input{pstex/geometry.pstex_t}
\end{center}
\mycaption{fig_geometry}{%
  (a) The tetrahedra sharing a common triangle (`exploded'
  drawing). The solid polygon denotes the corresponding dual face. A
  pair of $S^3$-variables (dots) is associated to each
  tetrahedron. The interaction term for this triangle depends on some
  characters ($\chi$) calculated for each pair of neighbouring
  tetrahedra (pairs of solid lines). However, each tetrahedron
  contributes with different $S^3$-values $(+)$ or $(-)$ depending on
  the neighbour. There is therefore some parallel transport associated to
  the tetrahedra (dashed lines). (b) The five tetrahedra in the
  boundary of a given four-simplex. This diagram visualizes all
  characters that are calculated for the weights associated to the
  various triangles. Each pair of lines belongs to one of the 10
  triangles involved in the four-simplex and coincides with a pair of
  lines in diagram~(a) when (a) is drawn for that triangle.}
\end{figure}

In this section, we restrict ourselves to the case in which the
interactions are local, \ie\ to the situation in which the edge
amplitude factorizes according to~\eqref{eq_edgefactor}. We consider a
$2$-complex $(V,E,F)$ that is dual to a given triangulation of a
four-manifold. For more general $2$-complexes dual to
a generic cellular decomposition, the interpretation has to be
generalized accordingly.

Using the identification $\SU(2)\cong S^3$, the partition
function~\eqref{eq_bcdual} suggests the following geometric
interpretation. The variables of the model are the pairs of
$S^3$-values associated to the tetrahedra of the triangulation. An
oriented tetrahedron, embedded in $\R^4$, spans an oriented hyperplane
which can be parameterized by its (normalized) normal vector, \ie\ by
a point in $S^3\subset\R^4$.

The partition function contains an interaction term
$f(g_1^{(+)},g_1^{(-)},\ldots)$ for each triangle. The interaction
depends on the $S^3$-values for all tetrahedra that contain the
triangle in their boundaries. Recall that the characters of $\SU(2)$
are given by
\begin{equation}
   \chi^{(j)}(g)=\frac{\sin((2j+1)\phi)}{\sin(\phi)},
\end{equation}
where $\phi$ denotes the polar angle on $S^3\cong\SU(2)$. The
interaction therefore depends only on the relative angles between the
normal vectors of each pair of neighbouring tetrahedra, the so-called
\emph{dihedral angles}. Two tetrahedra are neighbours with respect to
a given triangle if they are contained in the boundary of the same
$4$-simplex (this is the same as saying that on the dual $2$-complex,
two neighbouring edges contain a common vertex, but expressed in the
language of the original triangulation).

However, there is the complication that each tetrahedron carries two
$S^3$-variables, $g_e^{(+)}$ and $g^{(-)}_e$. For a given triangle,
each of these values is involved in the interaction term: the first
variable, $g_e^{(+)}$, in association with one neighbour and the
second variable, $g_e^{(-)}$, with the other one, see
Figure~\ref{fig_geometry}(a). The tetrahedron dual to $e$ therefore
contributes with different $S^3$-values $g_e^{(+)}$ and $g_e^{(-)}$
depending on the four-simplex in which the interaction takes place. It
is thus plausible to conjecture that there is some parallel transport
associated to the tetrahedra which contains the information of how
$g_e^{(-)}$ is rotated to $g_e^{(+)}$ when we transit from one
four-simplex along a tetrahedron to a neighbouring four-simplex. The
obvious candidate for the parallel transport is the transitive
$\SO(4)$-action on $S^3$ for which the $\SO(4)$ element is, however,
only defined up to an element in the stabilizer of a given point on
$S^3$. In Section~\ref{sect_bcgauge}, we reconstruct the $\SO(4)$
parallel transport, and our intuition about the ambiguity due to the
stabilizer is shown to be correct.

The weight~\eqref{eq_bcweight} resembles the character decomposition
of a function on $\SU(2)\times\SU(2)$ where the character
decomposition can be understood as a generalization of Fourier
expansion. Because of the geometric picture, the $S^3$-variables have
a natural interpretation\footnote{This interpretation was suggested by
D.~Oriti. Further work on the geometric interpretation at the
classical and at the quantum level is in progress~\cite{Or01b}.} as
variables conjugate to the representations labelling the faces ---
conjugate in the sense of the quantum mechanical uncertainty relation.

If we suppose that the partition function can be interpreted as the
trace over a suitable Hilbert space, there are two pictures for the
Barrett--Crane model in analogy to position and momentum basis in
Quantum Mechanics: In the original formulation, the partition function
is a sum over all possible quantized areas associated to the triangles
where the areas are determined by the balanced irreducible
representation of $\SO(4)$~\cite{Ba98a,BaBa00}. If we are in a quantum
state with definite areas of the triangles, the directions of the
tetrahedra in $\R^4$ which are represented by the $S^3$-variables and
the corresponding dihedral angles, would be maximally uncertain.

In addition to this `area basis' we also have a `direction basis'
which is given by the dual partition sum. The dual partition sum is
the integral over all possible $S^3$-directions of the tetrahedra. In
a quantum state with definite directions, however, we would expect a
maximal uncertainty of the quantized areas assigned to the triangles.

\subsection{The gauge theory picture}
\label{sect_bcgauge}

We have conjectured above that it is possible to find the parallel
transports of an $\SO(4)$ connection which, for each edge $e\in E$,
map the first $S^3$-variable $g_e^{(-)}$ to the second one
$g_e^{(+)}$. This would mean that the Barrett--Crane model has some
properties of an $\SO(4)$ gauge theory and therefore enjoys the
corresponding local symmetry. A suitable notion of local $\SO(4)$
symmetry is the invariance of the Boltzmann weight under a
transformation which takes an arbitrary $\SO(4)$ element for each
vertex $v\in V$ and transforms each edge variable
$(g_e^{(+)},g_e^{(-)})$ depending on the group elements that have been
chosen for the two vertices.

\begin{proposition}
Let $h\colon V\to\SO(4), v\mapsto (h_v,\tilde h_v)$ define a
generating function. Here $(h_v,\tilde h_v)\in\SU(2)\times\SU(2)$
represents a class of $\SO(4)\cong(\SU(2)\times\SU(2))/\Z_2$. The
Barrett--Crane model~\eqref{eq_bcdual} is invariant under the local
transformations,
\begin{mathletters}
\begin{eqnarray}
  g_e^{(+)}\mapsto h_{\del_+e}\cdot g_e^{(+)}\cdot\tilde h_{\del_+e}^{-1},\\
  g_e^{(-)}\mapsto h_{\del_-e}\cdot g_e^{(-)}\cdot\tilde h_{\del_-e}^{-1}.
\end{eqnarray}%
\end{mathletters}%
By Proposition~\ref{prop_su2action}, this transformation involves the
action of $\SO(4)$ on $S^3$ for each of the $S^3$-variables at the
given edge.
\end{proposition}

\begin{proof}
For a given face $f\in F$ and a vertex $v\in f_0$, the transformation
reads
\begin{mathletters}
\begin{eqnarray}
  g_{v_+(f)}^{(+)}\mapsto h_v\cdot g_{v_+(f)}^{(+)}\cdot\tilde h_v^{-1},\\
  g_{v_-(f)}^{(-)}\mapsto h_v\cdot g_{v_-(f)}^{(-)}\cdot\tilde h_v^{-1},
\end{eqnarray}
\end{mathletters}%
so that $\chi^{(j)}(g_{v_+(f)}^{(+)}\cdot{g_{v_-(f)}^{(-)}}^{-1})$ is
invariant.   
\end{proof}

It is now possible to reconstruct $\SO(4)$ parallel transports from
the pairs of $S^3$-values. Therefore we recall that the action of
$\SO(4)$ on $S^3$ given in Proposition~\ref{prop_su2action} is
transitive, \ie\ for two points $g_1,g_2\in S^3$, there exists a
$(g_1,g_2)\in\SO(4)$ such that
\begin{mathletters}
\begin{equation}
  (g_1,g_2)\ast g_2 = g_1.
\end{equation}%
Of course, we cannot see elements in the stabilizer of
$g_2$. Therefore for any $h\in\SO(3)$, the action is,
\begin{equation}
\label{eq_stab}
  ((g_1,g_2)\cdot \imath_{g_2}(h))\ast g_2 =
  ((g_1,g_2)\cdot (g_2hg_2^{-1},h))\ast g_2 = g_1.
\end{equation}%
\end{mathletters}%

We wish to start with an $\SO(4)$ gauge theory and then impose
the condition that elements from these stabilizers are irrelevant. In
other words, we wish to describe an $\SO(4)$ gauge theory of which only
the action of the parallel transport on $S^3$ is physically relevant.

Technically, the key idea is to consider a representation function of
$\SO(4)$ whose argument is the $\SO(4)$ element $(g_1,g_2)$ and to
average over the stabilizer as parameterized in~\eqref{eq_stab}. The
result of the averaging is indeed a product of two spherical functions
on $S^3\cong\SU(2)$ with arguments $g_2$ and $g_1$, respectively.

\begin{lemma}
\label{lemma_simpleproject}
Let $t^{(j{j^\prime}^\ast)}_{mn,pq}\in\Calg(\SO(4))$ and
$(g_1,g_2)\in\SO(4)$. Then
\begin{equation}
\label{eq_simpleproject}
  \int_{\SO(3)}t^{(j{j^\prime}^\ast)}_{mn,pq}((g_1,g_2)\cdot \imath_{g_2}(h))\,dh
    =\frac{\delta_{jj^\prime}}{2j+1}t^{(j)}_{mn}(g_1)\cdot t^{(j^\ast)}_{pq}(g_2).
\end{equation}
\end{lemma}

\begin{remark}
\begin{myenumerate}
\item
  If we integrate~\eqref{eq_simpleproject} over $\SO(4)$, the right
  hand side of this equation factorizes into two $\SU(2)\cong S^3$
  integrals. 
\item
  The above lemma crucially makes use of the special relations among
  $S^3$, $\SU(2)$ and $\SO(4)$. In particular, in higher dimensional
  generalizations~\cite{FrKr99,FrKr00}, we would still have
  $\SO(N+1)/\SO(N)\cong S^N$, but the right hand side
  of~\eqref{eq_simpleproject} would no longer factorize.
\item
  The averaging over the stabilizer automatically singles out the
  balanced irreducible representations of $\SO(4)$ for which
  $j=j^\prime$. 
\item
  The map
\begin{equation}
  \Calg(\SO(4))\to\Calg(\SO(4)),t^{(j{j^\prime}^\ast)}_{mn,pq}(g_1,g_2)\mapsto
    \int_{\SO(3)}t^{(j{j^\prime}^\ast)}_{mn,pq}((g_1,g_2)\cdot\imath_{g_2}(h))\,dh,
\end{equation}
  is a projector.
\item
  In Lemma~\ref{lemma_simpleproject}, we can replace $\SO(4)$ by
  $\Spin(4)\cong\SU(2)\times\SU(2)$ in which case the stabilizer is
  $\SU(2)$. The lemma is still satisfied. In particular, each balanced
  irreducible representation of $\Spin(4)$ is also a representation of
  $\SO(4)$.
\end{myenumerate}
\end{remark}

Lemma~\ref{lemma_simpleproject} suggests that it should be possible to
start with an $\SO(4)$ lattice gauge theory and then to apply the
averaging over the stabilizer in order to restrict the partition
function to the balanced irreducible representations and at the same
time split the $\SO(4)$ integral of the Haar intertwiner into the pair
of independent $S^3$-integrals in the definition of the Barrett--Crane
intertwiner.

In the remainder of this section, we prove that the Barrett--Crane
model is indeed obtained from lattice $\SO(4)$ $BF$-theory by applying
Lemma~\ref{lemma_simpleproject} in a suitable manner to the integrand
of the partition function. We therefore confirm that the
Barrett--Crane model agrees with $\SO(4)$ $BF$-theory subject to
constraints~\cite{DPFr99} whose form we derive at the quantum level.

We wish to apply the averaging procedure of
Lemma~\ref{lemma_simpleproject} to the $\SO(4)$ Haar
intertwiner~\eqref{eq_haarintermatrix} in order to obtain the
Barrett--Crane intertwiner~\eqref{eq_bcintermatrix} together with the
balancing (\emph{simplicity}) condition $j=j^\prime$. Obviously, we
need one averaging per edge and per face attached to that edge, namely
one averaging for each representation function in the Haar
intertwiner.

We therefore introduce the \emph{exploded configurations} of lattice
gauge theory which provide several copies of each edge variable so
that we have separate variables attaining the same value for all faces
attached to a given edge.

\begin{definition}
Let $(V,E,F)$ be a finite $2$-complex with cyclic structure. 
\begin{myenumerate}
\item
  We define the \emph{explosion map}
\begin{equation}
  X\colon {\SO(4)}^E\mapsto {\SO(4)}^{E\times F},
    {\{g_e\}}_{e\in E}\mapsto {\{g_{(e,f)}\}}_{(e,f)\in E\times F},
\end{equation}
  so that $g_{(e,f)}=g_e$. The elements of the set
  ${\SO(4)}^{E\times F}$ are called the \emph{exploded
  configurations}. 
\item
  The \emph{simplicity constraint},
\begin{mathletters}
\label{eq_simplicity}
\begin{equation}
  P\colon\Calg({\SO(4)}^{E\times F})\to\Calg({\SO(4)}^{E\times F}),
\end{equation}%
  is defined by specifying its image on a basis of the tensor product
  $\Calg({\SO(4)}^{E\times F})\cong\bigotimes_{(e,f)\in E\times
  F}\Calg(\SO(4))$, namely for each $(e,f)\in E\times F$,
\begin{eqnarray}
  t^{(j{j^\prime}^\ast)}_{mn,pq}(g_{(e,f)}^{(+)},g_{(e,f)}^{(-)})
  &\mapsto& \hat v_j\,(2j+1)\int_{\SU(2)}t^{(j{j^\prime}^\ast)}_{mn,pq}
     \biggl((g_{(e,f)}^{(+)},g_{(e,f)}^{(-)})\cdot\imath_{g_{(e,f)}^{(-)}}(h)\biggr)\,dh\nn\\
  &=&\hat v_j\delta_{jj^\prime}\,t^{(j)}_{mn}(g_{(e,f)}^{(+)})\cdot t^{(j^\ast)}_{pq}(g_{(e,f)}^{(-)}).
\end{eqnarray}%
\end{mathletters}%
\item
  Given any local Boltzmann weight $w\colon\SO(4)\to\R$, the
  \emph{exploded Boltzmann weight} $\tilde W\colon{\SO(4)}^{E\times
  F}\to\R$, is defined as,
\begin{equation}
  \tilde W({\{g_{(e,f)}\}}_{(e,f)\in E\times F}) := \prod_{f\in F} 
    w(g_{(\del_1f,f)}^{\epsilon_1f}g_{(\del_2f,f)}^{\epsilon_2f}\cdots
      g_{(\del_{N(f)}f,f)}^{\epsilon_{N(f)}f}).
\end{equation}
\end{myenumerate}
\end{definition}

We have absorbed the edge amplitude in its factorized
form~\eqref{eq_edgefactor} into the definition of the simplicity
constraint~\eqref{eq_simplicity}.

The partition function of $\SO(4)$ lattice gauge theory can, of
course, be written in terms of the exploded configurations.

\begin{proposition}
Let $(V,E,F)$ be a finite $2$-complex with cyclic structure. The
partition function of $\SO(4)$ lattice gauge theory on $(V,E,F)$ with
Boltzmann weight $w\colon\SO(4)\to\R$ is equal to
\begin{equation}
  Z = \Bigl(\prod_{e\in E}\,\int_{\SO(4)}\,dg_e\Bigr)\,
    (\tilde W\circ X)({\{g_e\}}_{e\in E}).
\end{equation}
\end{proposition}

We can now state our theorem about the reformulation of the
Barrett--Crane model as an $\SO(4)$ lattice gauge theory that is
subject to the simplicity constraint.

\begin{theorem}[Gauge theory picture]
Let $(V,E,F)$ be a finite $2$-complex with cyclic structure. The
partition function of $\SO(4)$ lattice gauge theory on the exploded
configurations, subject to the simplicity
constraint~\eqref{eq_simplicity},
\begin{equation}
\label{eq_constrainbf}
  Z = \Bigl(\prod_{e\in E}\,\int_{\SO(4)}\,dg_e\Bigr)\,
        (P(\tilde W)\circ X)({\{g_e\}}_{e\in E}),
\end{equation}
agrees with the Barrett--Crane model~\eqref{eq_bcmodel} for face
amplitudes,
\begin{equation}
\label{eq_bccharexp}
  \hat w_j=\int_{\SO(4)}\overline{\chi^{(j,j^\ast)}(g)}w(g)\,dg,
\end{equation}
that are the coefficients of the $\SO(4)$ character expansion of the
Boltzmann weight for the balanced irreducible representations. The
edge amplitudes are of the factorized form~\eqref{eq_edgefactor}. In
particular, for $\SO(4)$ lattice $BF$-theory, \ie\ $w(g)=\delta(g)$,
we obtain the Barrett--Crane model with $\hat w_j={(2j+1)}^2$. For
edge amplitudes~\eqref{eq_edgefactor} with coefficients $\hat
v_j=1/(2j+1)$, the simplicity constraint forms a projector.
\end{theorem}

\begin{proof}
First we consider the Boltzmann weight $\tilde W$ on the exploded
configurations and carry out the duality transformation for lattice
gauge theory until the analogue of~\eqref{eq_step20} for the exploded
configurations has been reached. We move all integrals from the Haar
intertwiners to the left of the expression and obtain the following
integrand,
\begin{eqnarray}
  \tilde W({\{g_{(e,f)}\}}_{(e,f)\in E\times F})
  &=& \Bigl(\prod_{f\in F}\sum_{\ontop{j_f,j_f^\prime\in\frac{1}{2}\N_0}{j_f+j_f^\prime\in\N_0}}\Bigr)
      \Bigl(\prod_{f\in F}\alignidx{\hat w_{(j_f{j_f^\prime}^\ast)}}\Bigr)
      \Bigl(\prod_{f\in F}\prod_{v\in f_0}\sum_{n(f,v)=1}^{2j_f+1}\sum_{m(f,v)=1}^{2j_f^\prime+1}\Bigr)\nn\\
  &&\times\prod_{e\in E}\Bigl(
      \prod_{f\in e_+}t^{(j_f{j_f^\prime}^\ast)}_{n(f,\del_+e)m(f,\del_+e),n(f,\del_-e)m(f,\del_-e)}(g_{(e,f)})\nn\\
  &&\qquad\times
      \prod_{f\in e_-}t^{({j_f^\ast}j_f^\prime)}_{n(f,\del_+e)m(f,\del_+e),n(f,\del_-e)m(f,\del_-e)}(g_{(e,f)})\Bigr).
\end{eqnarray}
Applying the simplicity constraint~\eqref{eq_simplicity}, we find
\begin{eqnarray}
  &&(P(\tilde W))({\{g_{(e,f)}\}}_{(e,f)\in E\times F})\nn\\
  &=& \Bigl(\prod_{f\in F}\sum_{j_f\in\frac{1}{2}\N_0}\Bigr)\,
      \Bigl(\prod_{f\in F}\alignidx{\hat w_{(j_fj_f^\ast)}}\Bigr)\,
      \Bigl(\prod_{e\in E}\prod_{f\in\delta e}\hat v_{j_f}\Bigr)\,
      \Bigl(\prod_{f\in F}\prod_{v\in f_0}\sum_{n(f,v)=1}^{2j_f+1}\sum_{m(f,v)=1}^{2j_f+1}\Bigr)\nn\\
  &&\times\prod_{e\in E}\Bigl(
      \prod_{f\in e_+}t^{(j_f)}_{n(f,\del_+e)m(f,\del_+e)}(g_{(e,f)}^{(+)})   
                      \overline{t^{(j_f)}_{n(f,\del_-e)m(f,\del_-e)}(g_{(e,f)}^{(-)})}\Bigr.\nn\\
  &&\qquad\Bigl.
      \prod_{f\in e_-}\overline{t^{(j_f)}_{n(f,\del_+e)m(f,\del_+e)}(g_{(e,f)}^{(+)})}
                      t^{(j_f)}_{n(f,\del_-e)m(f,\del_-e)}(g_{(e,f)}^{(-)})\Bigr),
\end{eqnarray}
where we write $g_{(e,f)}=(g_{(e,f)}^{(+)},g_{(e,f)}^{(-)})$ for the
$\SO(4)$ variables of the exploded configurations. Setting
$g_{(e,f)}=g_e$, we can integrate over $\SO(4)$ for each edge and
obtain two independent $\SU(2)$ integrals over $g_e^{(+)}$ and
$g_e^{(-)}$ per edge. The result agrees with~\eqref{eq_step12}.
\end{proof}

\begin{remark}
\begin{myenumerate}
\item
  In a similar way, we could implement more general edge amplitudes
  into the simplicity constraint even if they do not
  factorize. However, the condition that the simplicity constraint
  forms a projector, singles out a particular edge amplitude. In this
  case the amplitude factorizes (equation~\eqref{eq_edgefactor}), and
  $\hat v_j=1/(2j+1)$.
\item
  In the entire discussion, we can replace $\SO(4)$ by
  $\Spin(4)\cong\SU(2)\times\SU(2)$ and at the same time $\SO(3)$ by
  $\SU(2)$. Passing to the covering does not change any result.
\item
  The Boltzmann weight~\eqref{eq_bcweight} in the dual formulation of
  the Barrett--Crane model is obtained from the Boltzmann weight of
  lattice gauge theory by applying the simplicity
  constraint. This means in particular that
\begin{eqnarray}
  &&f(g_1^{(+)},g_1^{(-)},g_2^{(+)},g_2^{(-)},\ldots,g_k^{(+)},g_k^{(-)})\\
  &=& \int_{\SO(3)\times\cdots\times\SO(3)}
  w\biggl((g_1^{(+)},g_1^{(-)})\cdot\imath_{g_1^{(-)}}(h_1)\cdots
    (g_k^{(+)},g_k^{(-)})\cdot\imath_{g_k^{(-)}}(h_k)\biggr)\,dh_1\cdots dh_k,\nn
\end{eqnarray}
  where $(g_j^{(+)},g_j^{(-)})\in\SU(2)\times\SU(2)$ denote
  representatives in $\SO(4)\cong(\SU(2)\times\SU(2))/\Z_2$.
\end{myenumerate}
\end{remark}

It is no coincidence that the derivation of the group theory picture
in the present section closely resembles the reversed direction of the
construction of Oriti and Williams~\cite{OrWi01}. Our result which was
motivated by the idea that there is an $\SO(4)$ parallel transport
associated to the tetrahedra (dual edges), turns out to agree with a
`trivial gluing' in~\cite{OrWi01}. In~\cite{OrWi01}, a certain
prescription for the gluing of four-simplices is employed in order to
extend the geometrical constraints on a single
four-simplex~\cite{BaCr98} to the entire triangulation. Different ways
of imposing the constraints in the gluing procedure would give rise to
different edge amplitudes~\cite{Or01b}. In our formulation, the edge
amplitude $\hat v_j$ in~\eqref{eq_simplicity} looks a bit
artificial. For our choice of the constraint, only one amplitude seems
to be natural, namely the case in which $P$ is a projector. In order
to implement the constraints differently in the gluing procedure, one
would have to change the constraint $P$ and perhaps also the
combinatorics of the explosion map and would then recover a preferred
edge amplitude from the requirement that $P$ forms a projector. The
results of~\cite{OrWi01} demonstrate that there exist alternative
constraint maps $P$ that have a particular geometric interpretation
and that give rise to different edge amplitudes. Further work on these
matters is in progress~\cite{Or01b}.

\subsection{Regularizing the Boltzmann weight}
\label{sect_bcregular}

The reformulation of the Barrett--Crane model in terms of continuous
variables presented in Theorem~\ref{thm_bcdual}, offers a framework in
which one can study the ground state of the model and small
fluctuations around it, and which is also well suited for numerical
studies. In this section, we sketch some central ideas and also
comment on the question of which amplitudes $\hat w_j$ and
$\sym{A}^{(e)}$ are admissible in the partition
function~\eqref{eq_bcmodel}.

The amplitudes $\hat w_j$ and $\sym{A}^{(e)}$ are not fixed from
geometric considerations~\cite{BaCr98}. The simplest choice is,
\begin{equation}
  \hat w_j=\dim\alignidx{V_j\otimes V_j^\ast}={(2j+1)}^2,\qquad
  \sym{A}^{(e)}=1.
\end{equation}
In this case the Boltzmann weight~\eqref{eq_bcweight} behaves
similarly to a $\delta$-function whose arguments are the products
$g_{\ell+1}^{(+)}\cdot{g_\ell^{(-)}}^{-1}$. If the two $S^3$-values
of each pair agree, the weight~\eqref{eq_bcweight} diverges because
$\lim_{\phi\to0}\chi^{(j)}(\phi)=2j+1$. On the other hand, a simple
numerical calculation suggests that the weight is very small (\ie\
grows with $j$ only much more slowly than the peak at $\phi=0$)
whenever there is a pair of $S^3$-variables whose values do not agree.

In order to obtain a first intuition, consider
Figure~\ref{fig_geometry} and assume that the Boltzmann weight
vanishes whenever two $S^3$-variables (dots) that form a pair in one
of the characters (pair of solid lines) are not aligned. According to
the structure of double lines in Figure~\ref{fig_geometry}(b), there
exists a vanishing weight unless all five $S^3$-values agree. The
four-simplices would therefore be rigid, and the hyperplanes of all
five tetrahedra in the boundary of a given four-simplex would have to
be aligned. This is geometrically a very degenerate configuration in
which the four-simplex has zero four-volume. In addition, there is no
relation between the $S^3$-values assigned to different
four-simplices. If we transit from one four-simplex along a
tetrahedron to a neighbouring four-simplex and imagine that the change
in $S^3$-value is given by a parallel transport along the (edge dual
to the) tetrahedron, then there is no restriction on that parallel
transport.

However, these arguments are not quite correct because we did not
precisely quantify the value of the total Boltzmann weight if we
multiply very large (if $\phi=0$) and very small (if some other
$\phi\neq0$) numbers. We therefore have to regularize the weight
function. This can be done, for example, by a non-trivial edge
amplitude $\sym{A}^{(e)}$ or by introducing a one parameter family of
functions that tends to the degenerate face amplitude in a suitable
`limit'. Obvious choices of amplitudes are the following.

\begin{lemma}
Let $\hat w_j\in\C$ denote the face amplitudes, let $\hat v_j\in\C$
define the factorized edge amplitude~\eqref{eq_edgefactor}, and 
\begin{equation}
\label{eq_series}
  f_k(\phi_0,\ldots,\phi_k)=\sum_{j\in\frac{1}{2}\N_0}
    \alignidx{\hat w_j\hat v_j^k}\prod_{\ell=1}^k
      \frac{\sin((2j+1)\phi_\ell)}{\sin(\phi_\ell)}
\end{equation}
denote the Boltzmann weight~\eqref{eq_bcweight} for a $k$-gon,
expressed in terms of the dihedral angles $\phi_\ell\in[-\pi,\pi)$ on
$S^3$.
\begin{myenumerate}
\item
  For $\hat w_j={(2j+1)}^2$ and $\hat v_j=1$ or $\hat v_j=1/(2j+1)$,
  the series does not converge if all $\phi_j=0$.
\item
  If $\hat w_j={(2j+1)}^2$ and $\hat v_j=1/{(2j+1)}^2$, then $f_k$,
  $k\geq2$, converges pointwise for all $\phi_\ell$ because the
  characters are bounded, $|\sin(n\phi)/\sin(\phi)|\leq n$ for all
  $\phi$.
\item
  If $\hat w_j={(2j+1)}^2\exp(-\frac{j(j+1)}{\beta})$ and $\hat v_j=1$
  or $\hat v_j=1/(2j+1)$, then $f_k$ converges pointwise for all
  $\phi_\ell$ and any $\beta>0$.
\end{myenumerate}
\end{lemma}

Furthermore, a simple numerical calculation suggests that $f_k$ is
strictly positive only in case (3) provided that $\hat v_j=1/(2j+1)$,
and that in the cases (2) and (3), the function $f_k$ has a unique
maximum when all $\phi_\ell=0$. Both choices (2) and (3) yield a
non-trivial shape of the function $f_k$. Therefore the four-simplices
will be not exactly rigid and degenerate, and there will be a
non-trivial interaction between the four-simplices mediated by the
Boltzmann weights for each triangle.

In order to obtain a well-defined partition function~\eqref{eq_bcdual},
it is sufficient that the integrand is an $L^2$ function of the
$\SU(2)$-variables. However, for numerical studies and also for an
intuitive understanding, it is useful to have a pointwise convergent
series~\eqref{eq_series} so that we can always compare the function
values for different angles $\phi_j$.

If we insist on a strictly positive and regular Boltzmann weight
function and on the projector property of the simplicity constraint,
choice~(3) with $\hat v_j=1/(2j+1)$ for which the $\hat w_j$ are given
by the character expansion coefficients~\eqref{eq_bccharexp} of the
$\SO(4)$ heat kernel action (for one of the Casimir
operators)~\cite{Ro92,MoMu94}, deserves to be studied in greater
detail. In this case, there exists a limit $\beta\to\infty$ in which
the Boltzmann weight tends to the degenerate weight $\hat
w_j={(2j+1)}^2$ of the Barrett--Crane model.

Choice~(3) therefore defines a generalized Barrett--Crane model with a
bare `inverse temperature' $\beta=1/g_0^2$ or with a bare `coupling
constant' $g_0$.  This regularized model provides a framework in which
one could study the classical ground state, \ie\ the configuration of
maximal probability, and small fluctuations around it. The Boltzmann
weight~\eqref{eq_bcweight} has a non-trivial shape, and it makes sense
to ask the question of whether the interaction terms for the triangles
can be responsible for long range correlations.  A perturbative
analysis would correspond to an asymptotic expansion around
$\beta\to\infty$ which is precisely the interesting case.

In the model with generic $\beta$, the configurations of the spin foam
formulation of the Barrett--Crane model correspond to the coefficients
of the strong coupling expansion, \ie\ of an expansion for small
$\beta$, similarly to the situation in lattice gauge
theory~\cite{OePf01,PfOe02}. At strong bare coupling, the
configurations for small representations dominate.

At this point, some warnings are appropriate. The reformulation of the
Barrett--Crane model~\eqref{eq_bcdual} might exhibit significant
non-perturbative effects. For all choices of amplitudes $\hat w_j$ and
$\hat v_j$ discussed above, the classical ground state, \ie\ the state
of maximal probability, is the configuration in which all
tetrahedra of a given four-simplex are aligned. This configuration is
geometrically degenerate and spans only a zero four-volume, so that
one would need quantum effects in order to regularize the degeneracy
and to provide a proper four-volume. This in turn means that the
typical configurations one would find in the path integral, are `far
away' from the classical ground state. Confinement in pure $\SU(3)$
gauge theory is a famous example for such a non-classical
behaviour. In this case, perturbation theory cannot be expected to
describe the quantum system correctly because it is an expansion
around the wrong vacuum.

The new parameter $\beta$ and maybe the freedom to choose different
edge amplitudes, also raise the question of whether and how we have to
renormalize the model. From the above considerations, one might
conjecture that the Barrett--Crane model had to be renormalized along
the lines of lattice gauge theory. 

The choice of a triangulation or of a dual $2$-complex for a given
smooth manifold would correspond to a choice of coordinates, and the
triangulation had to be `fine' enough, \ie\ had to contain
sufficiently many simplices, for the aspects we wish to study. An
interesting question is certainly whether there exist long range
correlations which would arise from local degrees of freedom of the
model and correspond in some sense to the Euclidean analogue of the
graviton. Long range is here referred to in a combinatorial sense,
\ie\ variables that are separated by many simplices should still be
correlated. Provided that such long range effective degrees of freedom
exist, several scenarios are conceivable.

First, we might have to tune or fine-tune $\beta$ or the edge
amplitude in order to obtain an infinite or at least a very large
correlation length. Alternatively the model could be, at least for
some range of parameters, in a Coulomb like phase and exhibit
correlations that decay with the reciprocal distance (again in the
combinatorial sense). This would provide a scenario in which there are
always long range correlations and which is largely insensitive to the
choice of parameters.

After one has finally understood the spectrum of correlation lengths
of the model in a combinatorial sense of `length', one would have to
reconstruct metric information from the data provided by the model so
that, say in a suitable `classical' limit, the combinatorial
correlations could be interpreted as correlation functions with
respect to a metric distance. It is an interesting question how a
fundamental length scale might arise in such an approach.

Finally, we mention that two versions of the Barrett--Crane model,
\cite{DPFr00} with normalized intertwiner, and~\cite{PeRo01,OrWi01},
have edge amplitudes that do not factorize. In this case, the dual
model~\eqref{eq_bcintermediate} has a non-local Boltzmann weight, and
the classical ground state is no longer obvious and might even have
better properties. This is a second possibility, besides quantum
effects, how one could avoid the degenerate ground state. A more
detailed study of the possible edge amplitudes of the dual model is in
preparation.

%
\section{Conclusion}
%
\label{sect_conclusion}

We have presented a reformulation of the Euclidean Barrett--Crane
model in terms of dual $S^3$-variables. The new variables have a
geometric interpretation and can be understood as variables conjugate
to the quantized areas.

The fact that the dual model can be phrased in terms of continuous
variables, prepares the ground for the application of a variety of
analytical and numerical techniques. Interesting are in particular the
question of how the degenerate classical ground state can be
regularized, maybe by quantum effects, maybe by a non-local Boltzmann
weight for the dual model or maybe only if we introduce a cosmological
constant or couple gravity to matter. Also important is the question
of which correlations are relevant in the dual model.

The possible choices of edge amplitudes and regularization parameters
$\beta$ raise the question of whether and how one has to renormalize
the theory. On the one hand, the model is not immediately
triangulation independent, so that some refinement or sum over
triangulations will be necessary. On the other hand, the result
(however this will be achieved) is expected to be background
independent. This requires an entirely new interpretation of the
renormalization procedure as opposed to what is familiar from the
context of lattice gauge theory and Statistical Mechanics. We are
confident that the reformulation presented here offers a framework in
which these questions can be approached.

\acknowledgements

The author is grateful to Emmanuel College, Cambridge, for a Research
Fellowship. I would like to thank John Barrett, M\'aty\'as Kar\'adi, Alan
Macfarlane, Robert Oeckl, Daniele Oriti, Nuno Rom\~ao, Ruth Williams
and Toby Wiseman for many valuable discussions, for comments on the
manuscript and on relevant literature.

\newcommand{\hpeprint}[1]{\texttt{#1}}%
\newcommand{\hpspires}[1]{}%

\end{document}